\documentclass[prd,aps,eqsecnum]{revtex4}
\usepackage{color}
\usepackage{epsfig}
\usepackage{amsfonts}

\usepackage[hypertex]{hyperref}
\usepackage{amsmath,amssymb,amsthm,graphicx,latexsym}
\usepackage{subfigure}
\usepackage{pstricks}
\usepackage{color}

\usepackage{mathrsfs}

\newcommand{\be}{\begin{equation}}
\newcommand{\ee}{\end{equation}}
\newcommand{\bea}{\begin{eqnarray}}
\newcommand{\eea}{\end{eqnarray}}
\newcommand{\bml}{\begin{subequations}}
\newcommand{\eml}{\end{subequations}}
\newcommand{\bfig}{\begin{figure}}
\newcommand{\efig}{\end{figure}}

\newcommand{\del}{\delta}

\begin{document}

\title{DBI Galileon inflation in background SUGRA}

\author{Sayantan Choudhury\footnote{Electronic address: {sayanphysicsisi@gmail.com}} ${}^{}$
and Supratik Pal\footnote{Electronic address: {supratik@isical.ac.in}} ${}^{}$}
\affiliation{Physics and Applied Mathematics Unit, Indian Statistical Institute, 203 B.T. Road, Kolkata 700 108, India \\
%
}

\date{\today}
\begin{abstract}
We introduce a model of {\it potential driven DBI Galileon inflation}  in background ${\cal N}$=1, ${\cal D}$=4 SUGRA.
 Starting from D4-$\bar{D4}$ brane-antibrane in the bulk ${\cal N}$=2, ${\cal D}$=5 SUGRA
 including quadratic Gauss-Bonnet corrections, we derive an effective ${\cal N}$=1, ${\cal D}$=4 SUGRA by dimensional reduction,
 that results in a Coleman-Weinberg type Galileon potential. We employ this potential in modeling inflation and in subsequent study of  
primordial quantum fluctuations for scalar and tensor modes.
Further, we estimate the major observable parameters
in both {\it de Sitter (DS)} and {\it beyond de Sitter (BDS)}
limits and confront them with recent observational data from WMAP7 by using the publicly available code CAMB.
\end{abstract}


\maketitle
\tableofcontents
\section{\bf INTRODUCTION}

In the very recent days a good number of theoretical physicists
 have devoted their attention to the development of consistent modified gravity theories which play analogous role as 
dark energy or the cosmological constant \cite{ratra}, \cite{caroll}. In higher-dimensional setups as in the
 case of {\it DGP model} \cite{rdvali},\cite{def} where self-acceleration
is sourced by a scalar field (the zero helicity mode of the 5D graviton), these types of Infra-Red (IR) modification of gravity \cite{nima}, \cite{gia} play 
a crucial role. Moreover, the  {\it  DGP model} replicates the general
relativistic features due to non-linear interactions via the well known {\it Vainshtein mechanism} \cite{vain,suji,babi}.
 Despite its profound success it has got some serious limitations
 \cite{suji}, \cite{royma}, which are resolved by introducing a dynamical field, {\it aka}, Galileon \cite{espo,nicolis,felice}
 arising on the brane from the bulk in the DGP setup. The cosmological consequences of the Galileon models have 
been studied to some extent in \cite{fabio,chow,kobayashi,ajtoll}. Very recently, a natural extension to the scenario has been brought forward by tagging Galileon with the 
good old DBI model \cite{guss}, \cite{edmu}, resulting in so-called ``DBI Galileon'' \cite{claudia,trodden,goon,shun,sebar},
 that has reflected a rich structure from four dimensional cosmological point
of view. However, in most of the physical situations, this type of {\it effective}
gravity theories are plagued with additional degrees of freedom which often results in unwanted debris like ghosts, Laplacian instabilities etc \cite{de,dec1,feli}. 
Very recently a nobel effort towards the supersymmetric extension of the DBI Galileon model and its inflationary signature are discussed in \cite{sazi} and \cite{gaziql}.  
In the present paper we introduce a single scalar field model described by the D3 DBI Galileon originated
 from D4-$\bar{D4}$ brane anti-brane setup in the background of five dimensional local version of the supersymmetric theory (supergravity). 
This prevents the framework from having extra degrees of freedom as
well as {\it Ostrogradski instabilities} \cite{ost},  resulting in a higher-order derivative scalar field theory free from any such unwanted instabilities.
Nevertheless, a consistent field theoretic derivation of the effective potential commonly used in the context of DBI Galileon cosmology has not been brought forth so far.
On top of that, it is imperative to point out that the SUGRA origin of D3 DBI Galileon is yet to be addressed. 
In this article we plan to address both of these issues explicitly by deriving the inflaton potential from our proposed framework of DBI Galileon. 
It turns out that the derived action includes, in certain limits, the decoupling limit of DGP model as well as some consistent theories of massive
gravity; and it also includes the ``K-mouflage'' \cite{zio}, \cite{zioa} and also ``G/KGB'' inflation \cite{kobayashi}, \cite{yama,kamada,vik}. 
 Moreover, 
in general appearance of non-vanishing frame functions 
in the 4D action expedites breakdown of shift symmetry. Without shift symmetry, it may happen that the theory is unstable
against large renormalization.
The background action chosen in our model preserves shift symmetry of the single scalar field which gives it a firm footing from phenomenological point of view as well.

The plan of the paper is as follows.
First we propose a fairly general framework by taking the full DBI action 
 in D4 brane in the background of bulk ${\cal N}$=2, ${\cal D}$=5 supergravity \cite{sayan1,sayan2,sayan6,nills} including the quadratic modification
in Einstein's Hilbert action via Gauss-Bonnet correction in the bulk coming from two loop correction in string theory \cite{sayan19}.
 Hence, using dimensional reduction technique, we derive the effective action for DBI Galileon in D3 brane 
in the background of ${\cal N}$=1, ${\cal D}$=4 supergravity
 induced by the  quadratic correction in the geometry sector and study cosmological inflationary scenario therefrom.
We next engage ourselves in studying quantum fluctuations, by employing second order perturbative action for scalar and tensor modes in {\it de Sitter (DS)} and
{\it beyond de Sitter (BDS)} limits, and hence  
  calculate the primordial power spectrum of the scalar and tensor modes, their running and other observable parameters.
 We further confront our model with observations by using the publicly available code CAMB \cite{camb}, and 
find them to fit well with latest observational data from WMAP7\cite{wmap7} and expected to fit fair well with upcoming data from PLANCK\cite{planck}.


\section{\bf  The Background Action }
Let us demonstrate briefly the construction of DBI Galileon starting from 
${\cal N}$=2,${\cal D}$=5 SUGRA along with Gauss Bonnet correction in D4 brane set up. The full five dimensional model is described by the following action
\be\label{totac}S^{(5)}_{Total}=S^{(5)}_{EH}+S^{(5)}_{GB}+S^{(5)}_{D4~brane}+S^{(5)}_{Bulk Sugra}\ee
where
\be\label{5eh}S^{(5)}_{EH}=\frac{1}{2\kappa^{2}_{5}}\int d^{5}x\sqrt{-g^{(5)}}\left[R_{(5)}-2\Lambda_{5}\right],\ee
\be\label{5gb}S^{(5)}_{GB}=\frac{\alpha_{(5)}}{2\kappa^{2}_{5}}\int d^{5}x\sqrt{-g^{(5)}}
\left[R^{ABCD(5)}R^{(5)}_{ABCD}-4R^{AB(5)}R^{(5)}_{AB}+R^{2}_{(5)}\right]\ee
where $\alpha_{(5)}$ and $\kappa_{(5)}$ represent Gauss-Bonnet coupling and 5D gravitational coupling strength respectively. Additionally, 
$\Lambda_{(5)}$ and $g^{(5)}$ represent the 5D cosmological constant and the determinant of the 5D metric explicitly mentioned in equation(\ref{metric}).
The D4 brane action decomposed into two parts as 
\be\label{d4}S^{(5)}_{D4~ brane}=S^{(5)}_{DBI}+S^{(5)}_{WZ},\ee
where the {\it DBI} action and the {\it Wess-Zumino} action are given by respectively
\be\label{dbi}S^{(5)}_{DBI}=-\frac{T_{(4)}}{2}\int d^{5}x \exp(-\Phi)\sqrt{-\left(\gamma^{(5)}+B^{(5)}+2\pi\alpha^{'}F^{(5)}\right)},\ee
\be\begin{array}{lllllll}\label{WZ}\displaystyle S^{(5)}_{WZ}=-\frac{T_{(4)}}{2}
\int \sum_{n=0,2,4} \hat{C_{n}}\wedge \exp\left(\hat{B}_{2}+2\pi\alpha^{'}F_{2}\right)|_{4~form}\\~~~~~~
\displaystyle=\frac{1}{2}\int d^{5}x\sqrt{-g^{(5)}}\left\{\epsilon^{ABCD}\left[\partial_{A}\Phi^{I}\partial_{B}\Phi^{J}
\left(\frac{C_{IJ}B_{KL}}{4T_{(4)}}\partial_{C}\Phi^{K}\partial_{D}\Phi^{L}+\frac{\pi\alpha^{'}C_{IJ}F_{CD}}{2}\right.
\right.\right.\\ \left.\left.\left.~~~~~~
\displaystyle+\frac{C_{0}}{8T_{(4)}}B_{IJ}B_{KL}\partial_{C}\Phi^{K}\partial_{D}\Phi^{L}+
\frac{\pi\alpha^{'}C_{0}}{2}B_{IJ}F_{CD}\right)+2\pi^{2}\alpha^{'2}T_{(4)}C_{0}F_{AB}F_{CD}-T_{(4)}\nu(\Phi)\right]\right\}\end{array}\ee
where $T_{(4)}$ is the D4 brane tension, $\alpha^{'}$ is the Regge Slope, $\exp(-\Phi)$
is the closed string dilaton and $C_{0}$ is the Axion. Here and through out the article hat denotes a pull-back onto the D4 brane so that $\gamma_{AB}$ 
is the 5D induced metric on the D4 brane explicitly defined in equation(\ref{ind}). Here $\gamma^{(5)}$, $B^{(5)}$ and $F^{(5)}$ represent the determinant
of the 5D induced metric ($\gamma_{AB}$) and the gauge fields ($B_{AB},F_{AB}$) respectively.
The gauge invariant combination of rank 2 field strength tensor, appearing in D4 brane, is ${\cal F}_{AB}
=B_{AB}+2\pi\alpha^{'}F_{AB}$ and $\{F_{2},B_{2}\}$ represents 2-form $U(1)$ gauge fields which have the only non-trivial components along compact direction. 
On the other hand
$C_{4}$ has components only along the non-compact spacetime dimensions.
In a general flux compactification all fluxes may be turned on as the Ramond-Ramond (RR) forms
$F_{n+1}=dC_{n}$ (along with their duals) with $n=0,2,4$ and the Neveu Schwarz-Neveu Schwarz (NS-NS) flux $H_{3}=dB_{2}$.
Additionally the D4 brane frame function is defined as:
 \be\label{frrr1}\nu(\Phi)=\left(\nu_{0}+\frac{\nu_{4}}{\Phi^{4}}\right)\ee which is originated 
from interaction between D4-$\bar{D4}$ brane in string theory. Here $\nu_{0}$ and $\nu_{4}$ represent the constants characterizing the
interaction strength between  D4-$\bar{D4}$ brane.

In eqn(\ref{totac}) ${\cal N}=2, {\cal D}=5$ bulk supergravity action can be written as \cite{sayan1,sayan2,sayan6}
\be\begin{array}{lllll}\label{as1} S^{(5)}_{Bulk~Sugra}=\frac{1}{2}\int d^{5}x\sqrt{-g_{(5)}}
e_{(5)}\left[-\frac{M^{3}_{5}R^{(5)}}{2}+\frac{i}{2}\bar{\Psi}_{i\tilde{m}}
\Gamma^{\tilde{m}\tilde{n}\tilde{q}}\nabla_{\tilde{n}}
\Psi^{i}_{\tilde{q}}-{S}_{IJ}F^{I}_{\tilde{m}\tilde{n}}F^{I\tilde{m}\tilde{n}}-\frac{1}{2}
g_{\mu\nu}(D_{\tilde{m}}\Phi^{\mu})(D^{\tilde{m}}\Phi^{\nu})\right]\\ \displaystyle
~~~~~~~~~~~~~~~~~~~~~~~~~~~~~~~~~~~~~~~~~~~~~~~~~~~~~~~~~~~~~~~~~~~~~~~~~~~~~~ + {\rm Fermionic} + {\rm Chern-Simons} + {\rm Pauli~~ mass},\end{array}\ee

The rank-2 tensor field ${S}_{I J}$, appearing in the kinetic terms 
of the gauge fields, is the
restriction of the metric of the 5
dimensional space on the 4 dimensional manifold
of the scalar fields given by 
\be
{S}_{IJ} =
 -2 C_{I J K}h^{K} +
 3 h_{I} h_{J}
\label{alph} \; \;,
\ee
where 
$\; 
h_{I} = C_{IJK}
h^{J}h^{K}=
{S}_{IJ} h^{J}
\;$
and $\;g_{xy}= h_{x}^{I}
h_{y}^{J} {S}_{IJ} \;$ 
is the metric of the
$4$-dimensional manifold ${\mathcal{M}}_{4}$. 
In these equations we use $h_{x}^{I}=-\sqrt{\frac{3}{2}}
h^{I},_{x} $ and   $h_{Ix} = \sqrt{\frac{3}{2}}
 h_{I},_{x}$ along with the following constraints
\be
h^{I} h_{I}=1, \quad
 h_{x}^{I} h_{I}=h^{I}
 h_{Ix}=0 \; \; .
\label{relations}
 \ee
Here $C_{IJK}$ are
constants symmetric in the three indices satisfies the cubic constraint relationship
$C_{IJK}h^{I}h^{J}h^{K}=1$.
With the parity assignments we have adopted, $h^0$ is even, while
$h^x=\Phi ^x$ are odd. Furthermore on the fixed points where the odd quantities vanish,
$h^0=1$. Analogous relations hold for the $h_I$'s. 
In this context the 5-dimensional coordinates
$X^{A}=(x^{\alpha},y)$, where $y$ parameterizes the extra
dimension compactified on the closed interval $[-\pi R,+\pi R]$
and $Z_{2}$ symmetry is imposed. We consider 5D {\it Yang Mills} SUGRA model which is described by the 
field content $\{e^{\tilde{m}}_{\tilde{\mu}},\Psi^{i}_{\tilde{\mu}}, A^{I}_{\tilde{\mu}}, \lambda^{ia}, \Phi^{x}\}$
where $\tilde{\mu}=\left(\mu,5\right)$ are curved and $\tilde{m}=\left(m,\dot{5}\right)$ are flat 5D indices with $\mu, m$ their
corresponding 4D indices. The remaining indices are $I=0,1,.....,n$, $a=1,2,.....,n$ and $x=1,2,....,n$. The SUGRA multiplet consists of
the f$\ddot{u}$nfbien $e^{\tilde{m}}_{\tilde{\mu}}$, two graviphoton $A^{0}_{\tilde{\mu}}$ and two gravitini $\Psi^{i}_{\tilde{\mu}}$,
where $i=1,2$ is the simplectic $SU(2)_{R}$ index. Moreover, there exists n vector multiplets, counting the {\it Yang Mills} fields ($A^{a}_{\tilde{\mu}}$).
 The spinor and the scalar fields included in the vector multiplets are collectively denoted by $\lambda^{ia}$, $\Phi^{x}$ respectively. 
The indices {\it a} and {\it x} are flat and curved indices respectively of the 5D manifold ${\cal M}$. It is important to mention here that the Chern-Simons terms can be gauged away
assuming cubic constraints and $Z_2$
symmetry. Now we consider full particle spectrum
, the $Z_{2}$ even fields $\{e^{m}_{\mu}, e^{\dot{5}}_{5}, \Psi^{1}_{\mu}, \Psi^{2}_{5}, A^{0}_{5},A^{a}_{\mu}, \lambda^{1a}\}$ and
the $Z_{2}$ odd fields $\{e^{\dot{5}}_{\mu}, e^{m}_{5}, \Psi^{2}_{\mu}, \Psi^{1}_{5}, A^{0}_{\mu}, A^{a}_{5}, \lambda^{2a}, \Phi^{x}\}$
propagates in the bulk. For computational purpose it is useful to define the five
dimensional generating function($G$) of supergravity in this setup as
\be\label{gen}G=-3\ln\left(\frac{T+T^{\dagger}}{\sqrt{2}}\right)+K(\Phi,\Phi^{\dagger}), \ee
 where the supergravity K$\ddot{a}$hler moduli fields are given by 
\be\label{moduliiiii}T=\frac{(e^{\dot{5}}_{5}-i\sqrt{\frac{2}{3}}A^{0}_{5})}{\sqrt{2}}\ee which 
is assumed to be stabilized under first approximation and 
$K(\Phi,\Phi^{\dagger})$ represents generalized K$\ddot{a}$hler function.

 Including the kinetic term of the five dimensional field
$\Phi$ and rearranging into a perfect square, the 5D bulk supergravity action can be expressed as
\be\label{modsug}
S\supset\frac{1}{2}\int d^{4}x\int^{+\pi R}_{-\pi
R}dy\sqrt{-g_{5}}e_{(4)}e^{5}_{\dot{5}}\left[g^{\alpha\beta}G_{M}^{N}(\partial_{\alpha}\Phi^{M})^{\dagger}(\partial_{\beta}\Phi_{N})
+\frac{1}{g_{55}}\left(\partial_{5}\Phi-\sqrt{V^{(5)}_{bulk}(G)})\right)^{2}\right],\ee
where the 5D potential described by
\be\label{fpotw}V^{(5)}_{bulk}(G)=\exp\left(\frac{G}{M^{2}}\right)\left[\left(\frac{\partial
W}{\partial \Phi_{M}}+\frac{\partial G}{\partial
\Phi_{M}}\frac{W}{M^{2}}\right)^{\dagger}(G_{M}^{N})^{-1}\left(\frac{\partial
W}{\partial \Phi^{N}}+\frac{\partial G}{\partial
\Phi^{N}}\frac{W}{M^{2}}\right)-3\frac{|W|^{2}}{M^{2}}\right]\ee
where $W$ physically represents the superpotential
in the context of ${\cal N}=2,{\cal D}=5$ supergravity theory and expressed in terms of the holomorphic combination of the fields $\Phi,\Phi^{\dagger},T$ and $T^{\dagger}$.
The field equations in presence of Gauss-Bonnet term can be expressed as
\be\label{eeq}G^{(5)}_{AB}+\alpha_{(5)}H^{(5)}_{AB}=8\pi G_{(5)}T^{(5)}_{AB}-\Lambda_{(5)}g^{(5)}_{AB},\ee
where the covariantly conserved Gauss-Bonnet tensor 
\be\label{uiyt}H^{(5)}_{AB}=2R^{(5)}_{ACDE}R_{B}^{CDE(5)}-4R_{ACBD}^{(5)}R^{CD(5)}
-4R_{AC}^{(5)}R_{B}^{C(5)}+2R^{(5)}R_{AB}^{(5)}-\frac{1}{2}g^{(5)}_{AB}
\left(R^{ABCD(5)}R^{(5)}_{ABCD}-4R^{AB(5)}R^{(5)}_{AB}+R^{2}_{(5)}\right)\ee
 which acts as a source term. It is 
useful to introduce the 5D metric in conformal form 
\be\label{metric}ds^{2}_{4+1}=g_{AB}dX^{A}dX^{B}
=\frac{1}{\sqrt{h(y)}}ds^{2}_{4}+\sqrt{h(y)}\tilde{G}(y)dy^{2}=\exp(2A(y))\left(ds^{2}_{4}+R^{2}\beta^{2}dy^{2}\right),\ee
with warp factor 
\be\label{tgvd}\exp(2A(y))=\frac{1}{\sqrt{h(y)}}=\frac{R^2}{b^2_{0}\beta^{2}}\tilde{G}(y)=\frac{b^{2}_{0}}{R^{2}\left(\exp(\beta y)
+\frac{\Lambda_{(5)}b^{4}_{0}}{24R^{2}}\exp(-\beta y)\right)}
\ee and $ds^{2}_{4}=g_{\alpha\beta}dx^{\alpha}dx^{\beta}$ is FLRW counterpart. In order to write down
explicitly the expression for D4 brane action, the induced metric can be shown as
\be\label{ind}\gamma_{CD}=\frac{1}{\sqrt{h(y)}}
\left(g_{AB}+h(y)\tilde{G}_{AB}\partial_{C}\Phi^{A}\partial_{D}\Phi^{B}\right).\ee

The 5D energy momentum tensor for the set up reads
\be\begin{array}{lllllll}\label{em1}T^{total (5)}_{\alpha\beta}=T^{bulk(5)}_{\alpha\beta}+T^{D4 brane(5)}_{\alpha\beta}=
G^{N}_{M}(\partial_{\alpha}\Phi^{M})^{\dagger}(\partial_{\beta}\Phi_{N})
-g_{\alpha\beta}\left[g^{\rho\sigma}(\partial_{\rho}\Phi^{M})^{\dagger}(\partial_{\sigma}\Phi_{N})G^{N}_{M}
+g^{55}\left(\partial_{5}\Phi-\sqrt{V^{(5)}_{bulk}(G)}\right)^{2}\right]\\~~~~~~~~~~~~~~~~~~~~~~~~~~~~~~~~~~~~~~~~~~~~~~~
+\left[K_{MN;X}\partial_{\alpha}\Phi^{M}\partial_{\beta}\Phi^{N}+Kg_{\alpha\beta}-2\nabla_{(\alpha}G(\Phi,X)\nabla_{\beta)}\Phi
+g_{\alpha\beta}\partial_{\lambda}G(\Phi,X)\partial^{\lambda}\Phi\right.\\ \left.~~~~~~~~~~~~~~~~~~~~~~~~~~~~~~~~~~~~~~~~~~~~
~~~~~~~~~~~~~~~~~~~~~~~~~~~~~~~~~~~~-G_{MN;X}(\Phi,X)\Box\Phi\partial_{\alpha}\Phi^{M}\partial_{\beta}\Phi^{N}\right],\end{array}\ee
\be\begin{array}{llllll}\label{em2}T^{total (5)}_{55}=T^{bulk (5)}_{55}+T^{D4 brane(5)}_{55}=\frac{1}{2}\left(\partial_{5}\Phi
-\sqrt{V^{(5)}_{bulk}(G)}\right)^{2}
-\frac{1}{2}g_{55}g^{\rho\sigma}G^{N}_{M}(\partial_{\rho}\Phi^{M})^{\dagger}(\partial_{\sigma}\Phi_{N})
\\~~~~~~~~~~~~~~~~~~~~~~~~~~~~~~~~~~~~~~~~~~~~~~~
+\left[K_{MN;X}\partial_{5}\Phi^{M}\partial_{5}\Phi^{N}+Kg_{55}-2\nabla_{(5}G(\Phi,X)\nabla_{5)}\Phi
+g_{55}\partial_{\lambda}G(\Phi,X)\partial^{\lambda}\Phi\right.\\ \left.~~~~~~~~~~~~~~~~~~~~~~~~~~~~~~~~~~~~~~~~~~~~
~~~~~~~~~~~~~~~~~~~~~~~~~~~~~~~~~~~~-G_{MN;X}(\Phi,X)\Box\Phi\partial_{5}\Phi^{M}\partial_{5}\Phi^{N}\right].\end{array}\ee
On the other hand, the {\it Klein-Gordon} equation of motion in 5D can be expressed as
\be\begin{array}{llllll}\label{eqmos}\partial_{5}
\left[\frac{e^{5}_{\dot{5}}\sqrt{g_{5}}}{g_{55}}\left(\partial_{5}\Phi-\sqrt{V^{(5)}_{bulk}(G)}\right)\right]
+\sum_{N} e^{5}_{\dot{5}}\left\{\partial_{\beta}\left[\sqrt{g_{5}}g^{\alpha\beta}G^{N}_{M}(\partial_{\alpha}\Phi^{M})\right]
-\frac{\sqrt{g_{5}}}{g_{55}}\partial_{N}\left(\sqrt{V^{(5)}_{bulk}(G)}\right)
\left(\partial_{5}\Phi-\sqrt{V^{(5)}_{bulk}(G)}\right)\right\}\\
K_{X}(\Phi,X)\Box^{(5)}\Phi^{}-K_{XX}(\Phi,X)\left(\nabla_{A}\nabla_{B}\Phi\right)\left(\nabla^{A}\Phi\nabla^{B}\Phi\right)
-2K_{\Phi X}(\Phi,X)X+K_{\Phi}(\Phi,X)-2\left(G_{\Phi}(\Phi,X)-G_{\Phi X}(\Phi,X)X\right)\\ 
+G_{X}(\Phi,X)\left[\left(\nabla_{A}\nabla_{B}\Phi\right)\left(\nabla^{A}\nabla^{B}\Phi\right)
-\left(\Box^{(5)}\Phi\right)^{2}+R^{(5)}_{AB}\nabla^{A}\Phi\nabla^{B}\Phi
\right]+2G_{\Phi X}(\Phi,X)\left(\nabla_{A}\nabla_{B}\Phi\right)\left(\nabla^{A}\nabla^{B}\Phi\right)
+2G_{\Phi\Phi}(\Phi,X)X\\-G_{XX}\left(\nabla^{A}\nabla^{B}\Phi-g^{AB}\Box^{(5)}\Phi\right)
\left(\nabla_{A}\nabla^{C}\Phi\right)\left(\nabla_{B}\Phi\right)\left(\nabla_{C}\Phi\right)=0.\end{array}\ee

Now using the scaling relations 
\be\begin{array}{llll}\displaystyle {\Phi}^{A}=\sqrt{T_{(4)}}\tilde{\Phi}^{A},~~~~ G_{AB}=\exp(-\Phi)\tilde{g}_{AB},~~~~  
b_{AB}=\frac{\sqrt{h(y)}}{T_{(4)}}B_{AB}\end{array}\ee the 5D action for D4 brane can be expressed in more convenient form as
\be\label{conv}S^{(5)}_{D4~brane}=\int d^{5}x \sqrt{-g^{(5)}}\left[K(\Phi,X)-G(\Phi,X)\Box^{(5)}\Phi\right],\ee
where 
\be\label{kl}K(\Phi,X)=-\frac{1}{2f(\Phi)}\left(\sqrt{\cal D}
-1\right)-\frac{V^{(5)}_{brane}(\Phi)}{2}\ee where the determinant can be expressed as 
\be \label{detdef}{\cal D}\simeq 1-2f(\Phi)G_{AB}X^{AB}+4f^{2}(\Phi)X_{A}^{[A}X_{B}^{B]}
-8f^{3}(\Phi)X_{A}^{[A}X^{B}_{B}X^{C]}_{C}+16 f^{4}(\Phi)X_{A}^{[A}X^{B}_{B}X^{C}_{C}X^{D]}_{D}\ee
which is expressed in terms of the kinetic term 
$X^{B}_{D}=-\frac{1}{2}G_{DA}\partial^{C}\Phi^{A}\partial_{C}\Phi^{B}$. The detailed calculation for the determinant 
is elaborately discussed in the Appendex D. In this context the 5D D'Alembertian Operator is defined as
\be\label{5ddal}\Box^{(5)}=\frac{1}{\sqrt{-g^{(5)}}}\partial_{A}\left(\sqrt{-g^{(5)}}g^{AC}\partial_{C}\right).\ee
Here we use the fact that no spatial direction along which the scalar fields are only time dependent
lead to $B^{\mu}_{\nu}=0$ and $F_{\mu\nu}=0$ in the background. Consequently {\it Maxwell's field equations} are 
unaffected in 4D after dimensional reduction. In this context the 5D D4 brane potential is given by \be\label{D4pot}V^{(5)}_{brane}(\Phi)=T_{(4)}\nu(\Phi)+\frac{1}{f(\Phi)},\ee
where 5D warped geometry motivated $Z_{2}$ symmetric frame function 
\be\label{ffn} f(\Phi)=\frac{\exp(\Phi)h(y)}{T_{(4)}}
\simeq\frac{1}{(f_{0}+f_{2}\Phi^{2}+f_{4}\Phi^{4})}\ee  is originated from higher dimensional 
field theory and the implicit D4 brane function defined as:
\be\label{implbr}G(\Phi,X)=\frac{g(\Phi)}{2(1-2f(\Phi)X)}\ee
with $g(\Phi)=g_{0}+g_{2}\Phi^{2}$. Here $g_{0}$ and $g_{2}$ are model dependent constants
characterizes the effects of possible interactions on the D4 brane.

\section{\bf Modeling Inflation from D3 DBI Galileon} 

The technical details of the dimensional reduction technique are elaborately discussed in the Appendix A which can generate an effective D3 DBI Galileon theory in 4D.
Now summing up all the contributions from eqn(\ref{ghu}), eqn(\ref{gbe4}), eqn(\ref{br1})
  and eqn(\ref{ast5}), the model for {\it D3 DBI Galileon} is described by the 
following effective action:
\be\begin{array}{lllllll}\label{model1}
  \displaystyle S^{(4)}_{Total}= \int d^{4}x \sqrt{-g^{(4)}}
\left[\hat{\tilde{K}}(\phi,\tilde{X})
-\tilde{G}(\phi,\tilde{X})\Box^{(4)}\phi+\tilde{l}_{1}R_{(4)}\right.\\ \left.
\displaystyle~~~~~~~~~~~~~~~~~~~~~~~~~~~~~~~~~~~~~~~~~~~~~+\tilde{l}_{4}\left({\cal C}(1)R^{\alpha\beta\gamma\delta(4)}R^{(4)}_{\alpha\beta\gamma\delta}
-4{\cal I}(2)R^{\alpha\beta(4)}R^{(4)}_{\alpha\beta}+{\cal A}(6)R^{2}_{(4)}\right)+\tilde{l}_{3}\right],\end{array}\ee
where
\be\begin{array}{lllll}\label{effcons} \hat{\tilde{K}}(\phi,\tilde{X})=\tilde{K}(\phi,\tilde{X})-2X\tilde{M}(T,T^{\dag})-
\tilde{Z}(T,T^{\dag})V^{(4)}_{bulk}(\phi),\\ \tilde{M}(T,T^{\dag})=\frac{M(T,T^{\dag})}{2\kappa^{2}_{(4)}},
\tilde{Z}(T,T^{\dag})=\frac{Z(T,T^{\dag})}{2\kappa^{2}_{(4)}},\\ \tilde{l}_{1}=\left\{\frac{1}{2\kappa^{2}_{(4)}}
\left[1+\frac{\alpha_{(4)}}{R^{2}\beta^{2}}\left(24{\cal I}(2)-24{\cal A}(9)-16{\cal A}(10)\right)
\right]-\frac{\alpha_{(4)}{\cal C}(2)}{\kappa^{2}_{(4)}R^{2}\beta^{2}}\right\},\\ \tilde{l}_{4}=\frac{\alpha_{(4)}}{2\kappa^{2}_{(4)}}, \\
\tilde{l}_{3}=\frac{1}{2\kappa^{2}_{(4)}}\left[\frac{\alpha_{(4)}}{R^{4}\beta^{4}}
\left(24{\cal C}(24)-144{\cal I}(4)-64{\cal A}(5)+144{\cal A}(7)+64{\cal A}(8)+192{\cal A}(11)\right)-\frac{3M^{3}_{5}\beta b^{6}_{0}}
{2\kappa^{2}_{(4)}M^{2}_{PL}R^{5}}{\cal I}(1)\right]\end{array}\ee
where $\alpha_{(4)}$ and $\kappa_{(4)}$ are effective 4D Gauss-Bonnet coupling and gravitational coupling strength.
Here $\tilde{X}$ represents the 4D kinetic term after dimensional reduction. The other constants appearing in equation(\ref{model1}) and equation(\ref{effcons})
are explicitly mentioned in the Appendix C. For clarity, in terms of the effective potential the total four dimension 
action for our set up can be rewritten as:
\be\begin{array}{lllllll}\label{model1cd}
  \displaystyle S^{(4)}_{Total}= \int d^{4}x \sqrt{-g^{(4)}}
\left[K(\phi,\tilde{X})-\tilde{G}(\phi,\tilde{X})\Box^{(4)}\phi-V(\phi)
\right.\\ \left.
\displaystyle~~~~~~~~~~~~~~~~~~~~~~~~~~~~~~~~~~~~~~~~~~~~~+\tilde{l}_{1}R_{(4)}+\tilde{l}_{4}\left({\cal C}(1)R^{\alpha\beta\gamma\delta(4)}R^{(4)}_{\alpha\beta\gamma\delta}
-4{\cal I}(2)R^{\alpha\beta(4)}R^{(4)}_{\alpha\beta}+{\cal A}(6)R^{2}_{(4)}\right)+\tilde{l}_{3}\right],\end{array}\ee

 where \be\begin{array}{lll}\label{nwq1}
           \displaystyle K(\phi,\tilde{X})=-\frac{\tilde{D}}{\tilde{f}(\phi)}\left[\sqrt{1-2Q\tilde{X}\tilde{f}}-Q_{1}\right]
-\tilde{C}_{5}\tilde{G}(\phi,\tilde{X})-2X\tilde{M}(T,T^{\dag})
          \end{array}\ee
 For details see Appendix A.
 The collective effect of equation(\ref{scaled}) and equation(\ref{pot1}) gives the total {\it D3
DBI Galileon} potential appearing in equation(\ref{model1cd}) as:
\be\begin{array}{lllllllll}\label{modelpot} 
\displaystyle V^{}(\phi)=\bar{Q}_{2}\tilde{D}V^{(4)}_{brane}+
\tilde{Z}(T,T^{\dag})V^{(4)}_{bulk}(\phi)\\ \displaystyle ~~~~~~~=\sum^{2}_{m=-2,m\neq-1}C_{2m}\phi^{2m},\end{array}\ee
where 
\be\begin{array}{llll}\label{vccvc}\displaystyle C_{0}=\left(T^{}_{3}\tilde{\nu}_{0}+\beta R {\cal I}(2)\tilde{f}_{0}+\tilde{Z}(T,T^{\dag}){\cal A}(13)v^4\right),\\
C_{-4}=T^{}_{3}\tilde{\nu}_{4},\\ C_{2}=\left(\beta R {\cal I}(2)\tilde{f}_{2}-gv^{2}\tilde{Z}(T,T^{\dag}){\cal A}(13)\right),\\
C_{4}=\left(\beta R {\cal I}(2)\tilde{f}_{4}+\frac{\tilde{Z}(T,T^{\dag}){\cal A}(13)g^{2}}{4}\right)\end{array}\ee are tree level 
constants. Now we want to see the effect of one-loop radiative correction to the derived potential. After doing proper analysis
throughout it comes out that the one-loop correction does not effect the superpotential due to the cancellation of all tadpole terms appearing in the theory.
 On the other hand one-loop radiative correction in the K$\ddot{a}$hler potential results in 
\be\begin{array}{llll}\label{bvcxv}
  \displaystyle  \delta {\cal K}^{1-loop}(\phi,\phi^{\dagger})
=\int^{\Lambda_{UV}}_{p}\frac{d^{4}p}{(2\pi)^{4}p^{2}}\left[\frac{1}{2}Tr\ln \hat{\cal K}(\phi,\phi^{\dagger})
+\frac{1}{2}Tr\ln\left(\hat{\cal K}(\phi,\phi^{\dagger})p^{2}-\hat{\cal W}^{\dagger}(\phi,\phi^{\dagger})\left(\hat{\cal K}(\phi,\phi^{\dagger})^{-1}\right)^{\dagger}
\hat{\cal W}(\phi,\phi^{\dagger})\right)\right]\\
~~~~~~~~~~~~~~~~~~~~\displaystyle =\frac{\Lambda^{2}_{UV}}{16\pi^{2}}\ln\left(det\left[\hat{\cal K}(\phi,\phi^{\dagger})\right]\right)
-\frac{1}{32\pi^{2}}Tr\left({\cal M}^{2}_{\phi}\left[\frac{{\cal M}^{2}_{\phi}}{\Lambda^{2}_{UV}}-1\right]\right)
   \end{array}\ee
where $\Lambda_{UV}$(=M~(Reduced Planck Mass)) is used as a UV cut-off of the theory appearing in the context of cut-off regularization. In this connection
the chiral mass matrix is given by \be\label{massmatrix}
{\cal M}^{2}_{\phi}=\hat{\cal K}^{-\frac{1}{2}}(\phi,\phi^{\dagger})\hat{\cal W}^{\dagger}(\phi,\phi^{\dagger})
\left(\hat{\cal K}(\phi,\phi^{\dagger})^{-1}\right)^{\dagger}\hat{\cal W}(\phi,\phi^{\dagger})\hat{\cal K}^{-\frac{1}{2}}(\phi,\phi^{\dagger}).\ee
Now including the contribution 
from one-loop radiative correction both from brane and bulk SUGRA, the renormalizable potential is as under
\be\begin{array}{lll}\label{loop}
V(\phi)=V_{tree}(\phi)+\del V_{1-loop}(\phi)\\~~~~~~~
\displaystyle=\underbrace{\sum^{2}_{m=-2,m\neq-1}C_{2m}\phi^{2m}}_{Tree-level~ contribution}
+
\underbrace{\lim_{\epsilon\rightarrow 0}\sum^{0}_{n=-2,n\neq-1}B_{2m}\left(\int^{\Lambda_{UV}=M}_{p}\frac{d^{4}p}
{(2\pi)^{4}}\frac{1}{\left(p^{2}-2C_{2}+i\epsilon\right)^{2}}\right)\phi^{2n}}_{One-loop~ correction~ in~ D3~ brane}
\\~~~~~~~~~~~~~~~~~~~~~~~~~~~~~~~~~~\displaystyle +\underbrace{\sum^{2}_{q=0}\frac{\phi^{2q}}{64\pi ^{2}}\left[ \Lambda^{4}_{UV} STr\left(
{\cal M}^{0}\right) \ln \left( \frac{\Lambda^{2}_{UV}}{\phi^{2}}\right)
+2\Lambda^{2}_{UV} STr\left( {\cal M}^{2}\right) + STr \left( {\cal M}^{4}\ln \left(
\frac{{\cal M}^{2}}{\Lambda^{2}_{UV}}\right) \right) \right]}_{One-loop~ correction~ in~ the~ bulk~ {\cal N}=1,~ {\cal D}=4 ~SUGRA}\\
~~~~~~~
\displaystyle=\sum^{2}_{m=-2,m\neq-1}C_{2m}\phi^{2m}+\sum^{0}_{n=-2,n\neq-1}\bar{B}_{2n}\ln\left(\frac{\phi}{M}\right)\phi^{2n}
+\sum^{2}_{q=0}A_{2q}\ln\left(\frac{\phi}{M}\right)
\phi^{2q}\\
\\~~~~~~~\displaystyle=\sum^{2}_{m=-2,m\neq-1}\left[1+D_{2m}\ln\left(\frac{\phi}{M}\right)\right]\phi^{2m},\end{array}\ee
where $D_{0}=0$, $D_{2m}=\frac{\bar{B}_{2m}+A_{2m}}{C_{2m}}$ and we have used the supertrace identity
\be\label{supertrace}
STr \left( {\cal M}^{\alpha}\right) \equiv \sum_{i}\left( -1\right) ^{2j_{i}}\left(
2j_{i}+1\right) m_{i}^{\alpha}.\ee

It is the Coleman Weinberg potential \cite{coleman}, provided the coupling constant satisfies the
Gellmann-Low equation \cite{ryder} in the context of Renormalization group. Here the first term in the eqn(\ref{loop})
 physically represents the energy scale of inflation ($\sqrt[4]{C_{0}}$). Here the four dimensional effective potential respect the 
{\it Galilean symmetry}: $\phi\rightarrow \phi+b_{\mu}x^{\mu}+c$ which taske care both shift and spacetime translational symmetry.

\begin{figure}[htb]
{\centerline{\includegraphics[width=12cm, height=8.5cm] {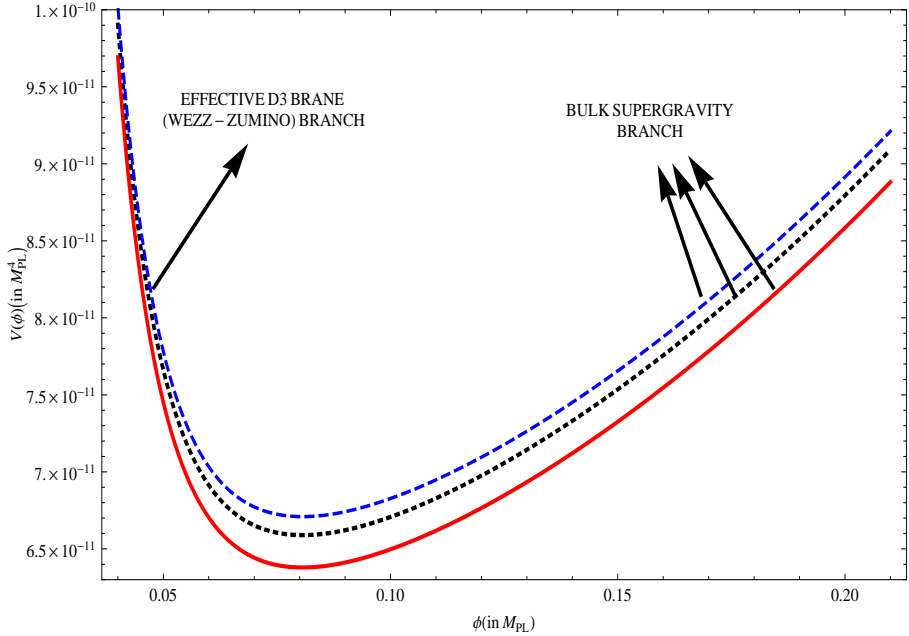}}}
\caption{Variation of one loop corrected potential($V(\phi)$) with inflaton field ($\phi$)} \label{figVr845}
\end{figure}

Figure (\ref{figVr845}) represents the inflaton potential for
different values of $C_{2m}$ and $D_{2m}$. From the observational constraints
the best fit model is given by the range: $5.67\times 10^{-11}M^{4}_{PL}<C_{0}<6\times 10^{-11}M^{4}_{PL}$,
 $1.01\times 10^{-16}M^{8}_{PL}<C_{-4}<2\times 10^{-16}M^{8}_{PL}$, $7.27\times 10^{-10}M^{2}_{PL}<C_{2}< 7.31\times 10^{-10}M^{2}_{PL}$,
$2.01\times 10^{-14}<C_{4}<2.45\times 10^{-14}$, $0.014 <D_{-4}<0.021$, $0.002<D_{2}< 0.012$  and $0.011<D_{4}<0.019$ so
that while doing numericals, we shall restrict ourselves to this
range. Here $M_{PL}=2.43\times 10^{18}GeV$ represents reduced 4D Planck mass. 
Consequently the energy scale of inflation has a window $0.658\times 10^{16}GeV<\sqrt[4]{C_{0}}<0.667\times 10^{16}GeV$ which precisely falls within the GUT scale.

 Hence using equation(\ref{model1}) the modified {\it Friedmann } and
{\it Klein-Gordon} equations can be expressed as:
\be\label{fr1}
H^{4}=\frac{\Lambda^{}_{(4)}+8\pi G_{(4)}V(\phi)}{\tilde{g}_{1}},\ee
\be\label{fr2}
\dot{\phi}^{2}\left(e_{2}(\phi)+9e_{3}(\phi)H^{2}\right)=\left\{V^{'}(\phi)+\tilde{C}_{5}\tilde{g}^{'}(\phi)k_{1}
-\frac{\tilde{D}\tilde{f}^{'}(\phi)}{\tilde{f}(\phi)}\left(1-Q_{1}\right)\right\},\ee
where $e_{2}(\phi)=\tilde{M}(T,T^{\dag})J_{\phi}+2\tilde{g}(\phi)
\tilde{f}^{'}(\phi)k_{1}k_{2}
+8\tilde{f}(\phi)\tilde{f}^{'}(\phi)\tilde{g}(\phi)k_{1}k^{2}_{2}
+2\tilde{g}^{'}(\phi)\tilde{f}(\phi)k_{1}k_{2}-g^{''}(\phi)k_{1}$ and 
$e_{3}(\phi)=2\tilde{C}_{4}\tilde{f}(\phi)\tilde{g}(\phi)k_{1}k_{2}$
provided $|e_{3}(\phi)|\gg|e_{2}(\phi)|$ in the slow-roll regime.
This has been
discussed in details in the Appendix B.
Here we have fixed the signature of $\dot{\phi}$ so that the scalar
field rolls down the potential. Additionally ghost instabilities are 
avoided provided the coefficient of $\dot{\phi}^{2}>0$. Consequently 
the potential dependent slow-roll parameters can be expressed as:
\bea \label{first} \epsilon_{V}
:&=&\frac{M^{2}_{PL}}{2}\left(\frac{V^{'}}{V}\right)^{2}\frac{1}{\sqrt{{\cal G}(\phi)V^{'}(\phi)}},
\\
\label{second}\eta_{V} :&=&
M^{2}_{PL}\left(\frac{V^{''}}{V}\right)\frac{1}{\sqrt{{\cal G}(\phi)V^{'}(\phi)}},
\\
\label{third} \xi_{V} :&=&
M^{4}_{PL}\left(\frac{V^{'}V^{'''}}{V^{2}}\right)\frac{1}{{\cal G}(\phi)V^{'}(\phi)},
\\
\label{fourth} \sigma_{V} :&=&
M^{6}_{PL}\left(\frac{(V^{'})^{2}V^{''''}}{V^{3}}\right)\frac{1}{\left({\cal G}(\phi)V^{'}(\phi)\right)^{\frac{3}{2}}},
\eea
where ${\cal G}(\phi)=\frac{16e_{3}(\phi)M^{2}_{PL}}{\tilde{g}_{1}V(\phi)}$. In this connection Galileon terms effectively flatten the potential
due to the presence of the {\it flattening factor} $\frac{1}{\sqrt{{\cal G}(\phi)V^{'}(\phi)}}\ll 1$. This implies that in the presence of Galileon
like derivative interaction slow-roll inflation can take place even if the potential is rather steep.

The number of e-foldings for {\it D3 DBI Galileon} can be expressed as 
\be\label{iop}
{\cal N}=\frac{1}{8M_{PL}}\int^{\phi_{f}}_{\phi_{i}}\frac{\sqrt[4]{{\cal G}(\phi)V^{'}(\phi)}}{\sqrt{\epsilon_{V}}}d\phi,\ee
where $\phi_{i}$ and $\phi_{f}$ are the corresponding values of
the inflaton field at the beginning and end of inflation.

\begin{figure}[h]
{\centerline{\includegraphics[width=10cm, height=8cm] {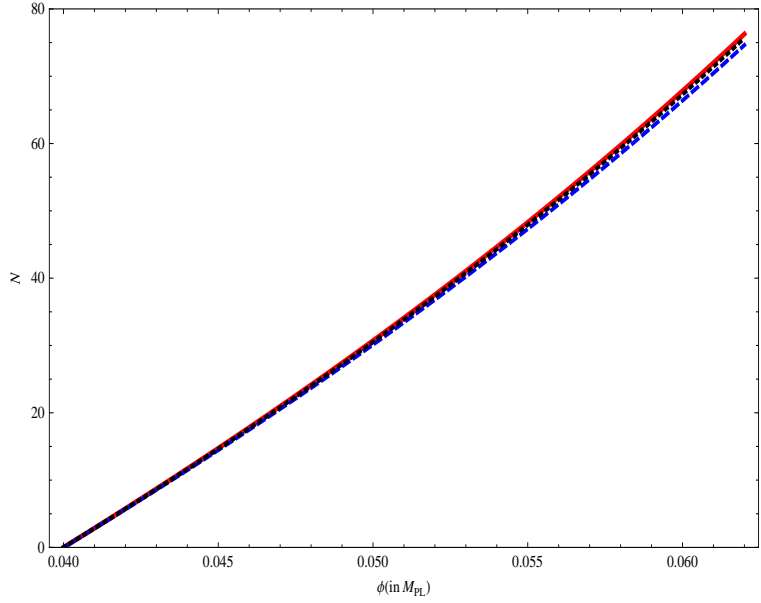}}}
\caption{Variation of the number of e-foldings ($N$)
 with inflaton field ($\phi$)} \label{fig:subfig1}
\end{figure}

\begin{figure}[ht]
\centering
\subfigure[]{
    \includegraphics[width=8cm,height=7cm] {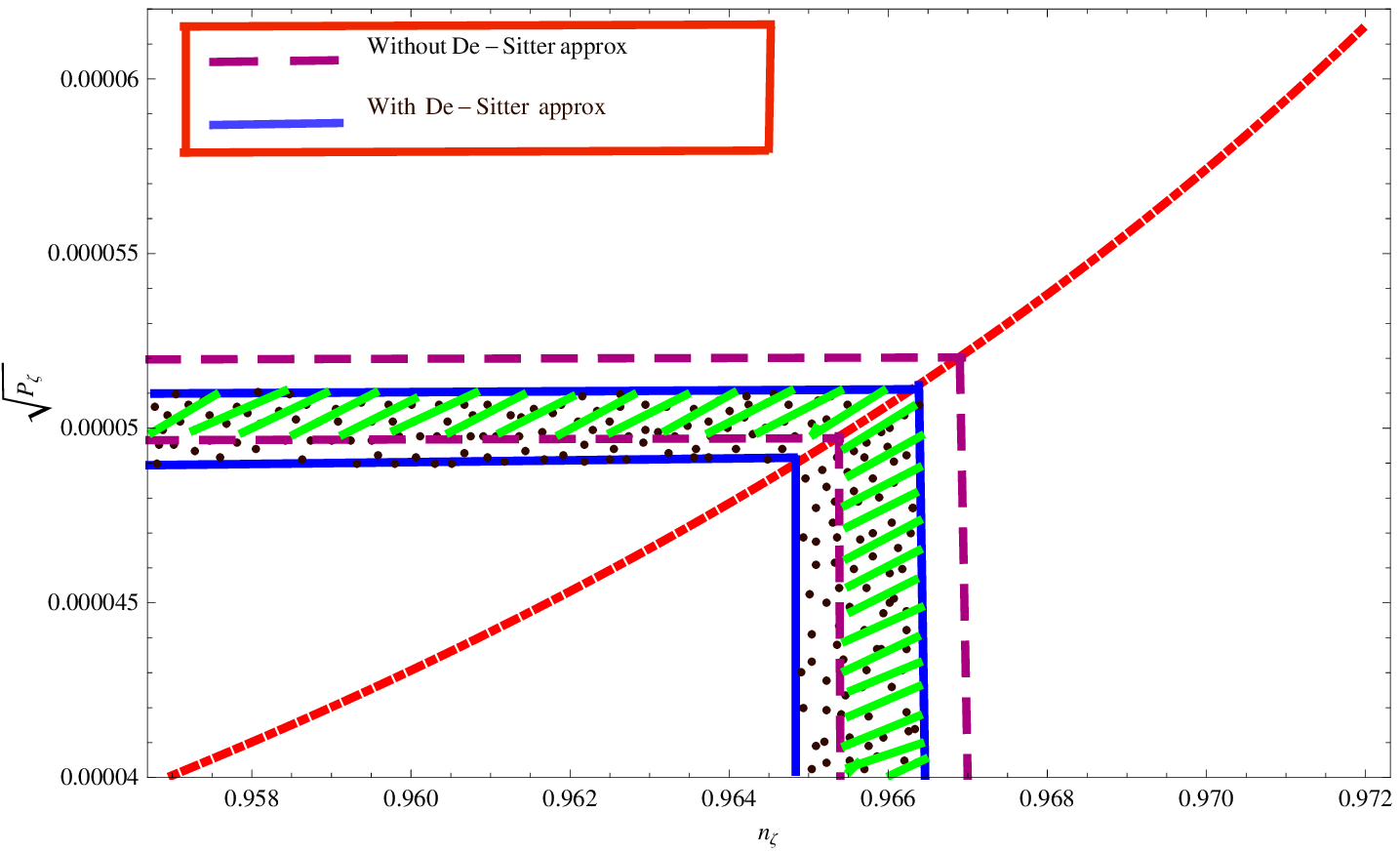}
    \label{fig:subfig2}
}
\subfigure[]{
    \includegraphics[width=8cm,height=7cm] {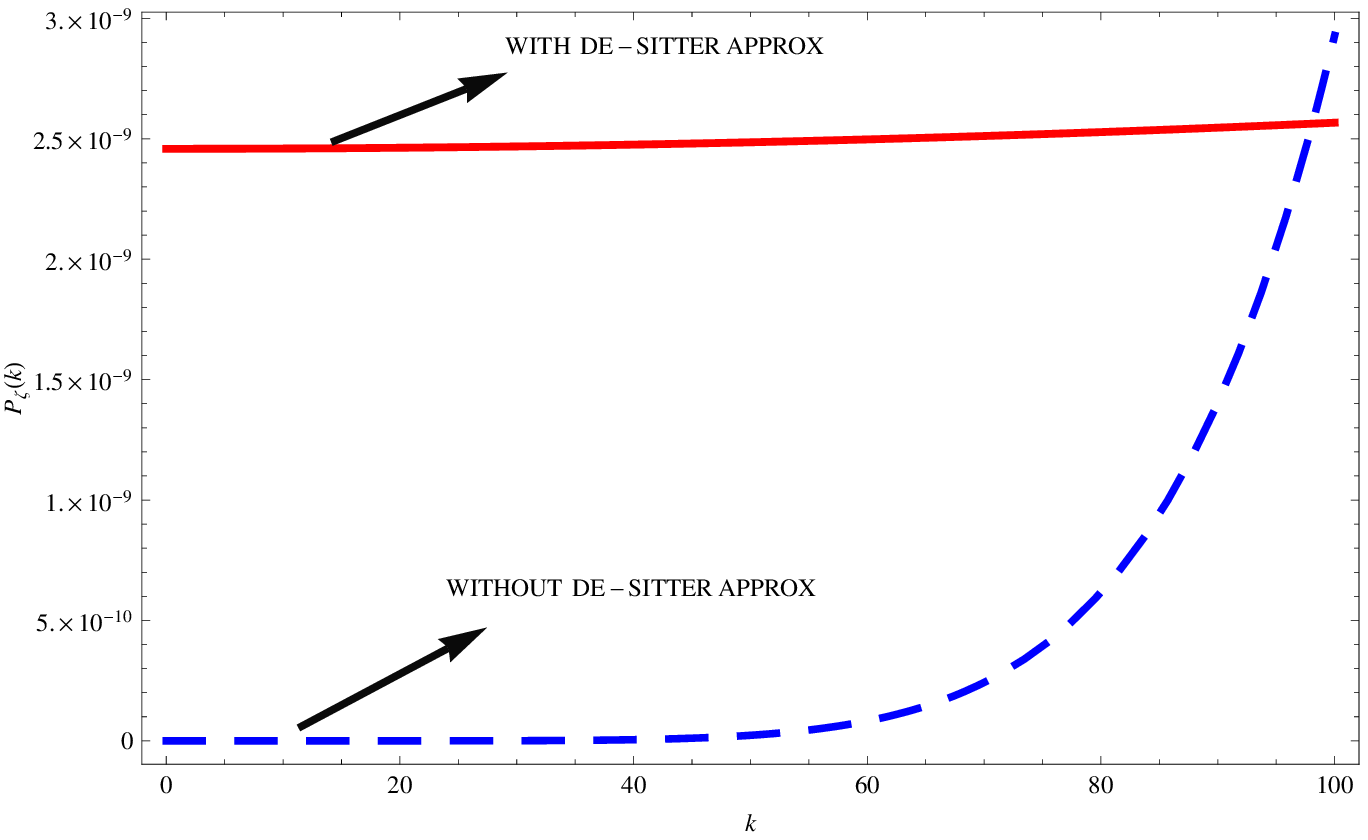}
    \label{fig:subfig3}
}
\subfigure[]{
    \includegraphics[width=8cm,height=7cm] {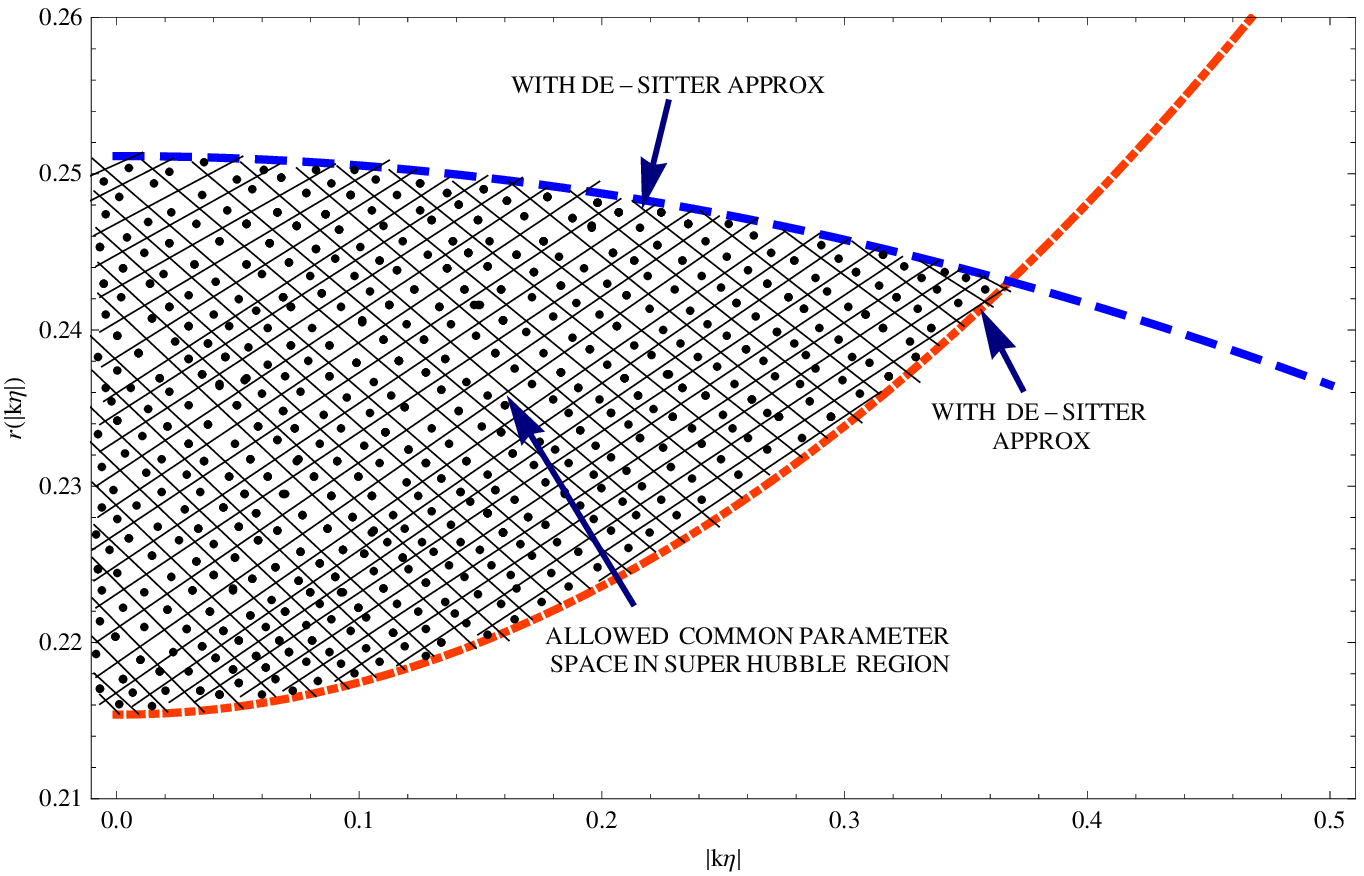}
    \label{fig:subfig4}
}
\caption[Optional caption for list of figures]{\subref{fig:subfig2}Parametric plot of the amplitude of the power spectrum ($\sqrt{{\cal P}_{\zeta}}$)
with scalar spectral index ($n_{\zeta}$),
\subref{fig:subfig3}Variation
 of the scale dependent scalar power spectrum 
(${\cal P}_{\zeta}(k)$) vs momentum scale k, \subref{fig:subfig4}Variation of the scale 
dependent tensor to scalar ratio ($r(k\eta)$) vs $|k\eta|$ for {\it DS} and {\it BDS} analysis.}
\label{fig:subfigureExample1}
\end{figure}

Figure(\ref{fig:subfig1}) represents a graphical behavior of number
of e-folding versus the inflaton field 
for different values of $C_{i}^{'}s$ and the most satisfactory point in
this context is the number of e-folding lies within the
observational window  $56<N<70$. The end of the inflation leads
to the extra constraint $V^{'}(\phi_{f})=\sqrt{V^{''}(\phi_{f})V(\phi_{f})}$.

\section{\bf Quantum Fluctuations and Observable Parameters}
Let us now engage ourselves in analyzing quantum fluctuation in
our model and its observational imprints via primordial spectra
generated from cosmological perturbation. To serve this purpose we start with the {\it ADM 
formalism} with the line element
\begin{equation}\label{metric2}
ds^{2}=-N^{2}dt^{2}+h_{ij}(N^{i}dt+dx^{i})
(N^{j}dt+dx^{j})\,,
\end{equation}
where $N$ and $N^i$ ($i=1, 2, 3$) are the lapse and shift functions, respectively.
In this context we consider scalar metric perturbations 
about the flat FLRW background. 
Here we expand the lapse $N$ and the shift vector
$N^{i}$, as $N=1+\alpha$ and $N_{i}=\partial_{i}\psi+\tilde{N}_{i}$, respectively. Here $\partial_{i}\psi$ is the irrotational part and 
$\tilde{N}_{i}$ be the incomressible vector part ($\tilde{N}_{i,i}=\partial_{i}\tilde{N}_{i}=0$).
These are actually non-dynamical Lagrange multipliers in the action, so that it is sufficient to know $N$ and $N^i$
up to first order. This implies their equation of motion is purely algebraic. 
To fix the time and spatial reparameterizaion we choose the uniform-field gauge 
with $\delta\phi=0$, which fixes the temporal component
of a gauge-transformation vector $\xi^{\mu}$. After that by fixing
the spatial part of $\xi^{\mu}$ we gauge away a field
$\varepsilon$ that appears as a form $\varepsilon_{,ij}$ inside $h_{ij}$. Consequently the metric on three dimensional constant time slice
can be expressed as $h_{ij}=a^{2}(t)e^{2{\zeta}}\delta_{ij}$. Plugging all of these informations in to equation(\ref{metric2})
we get:
\begin{equation}\label{eqb}
ds^{2}=-\left[ (1+\alpha)^{2}-a^{-2}(t) e^{-2{\zeta}}\left\{(\partial\psi)^{2}+\tilde{N}^{2}+2\tilde{N}.(\partial\psi)\right\}
\right]\, dt^{2}+2\left[\partial_{i}\psi+\tilde{N}_{i}\right]\, dt\, dx^{i}+a^2(t) e^{2{\zeta}}\delta_{ij}dx^{i}dx^{j},
\end{equation}
where we use a shorthand notation $(\partial\psi)^{2}=(\partial\psi_i)\,(\partial\psi_i)\equiv(\partial_{i}\psi)(\partial_{i}\psi)$. 
At linear level of the perturbation theory equation(\ref{eqb}) reuces to the following metric: 
\begin{equation}
ds^{2}=-(1+2\alpha)\, dt^{2}+2\left[\partial_{i}\psi+\tilde{N}_{i}\right]\, dt\, dx^{i}
+a^2(t)\, (1+2\zeta)\,\delta_{ij}dx^{i}dx^{j}\,.
\end{equation}

Expanding the four dimensional effective action stated in equation(\ref{model1cd}) up to second order, we get:
\begin{equation}
\label{eqxz}
S^{\zeta}_2=\int dt\,d^3x\,a^3
\left[ -3t_1 \dot{\zeta}^2 
+\frac{2w_1}{a^2} \dot{\zeta} \partial^2\psi
-\frac{t_{2}}{a^2} \alpha \partial^2\psi
-\frac{2 t_{1}}{a^2} \alpha \partial^2{\zeta}
+3 t_{2}\,\alpha\,\dot{\zeta}
+\frac13 t_{3} \alpha^2
+\frac{t_{4}}{a^2}\,\partial_i{\zeta}\,\partial_i{\zeta}\right]\,,
\end{equation}
where the effect of effective Gauss-Bonnet coupling and the DBI Galileon features 
are explicitly appearing in the co-efficients of the second order perturbative action as:
\begin{eqnarray}
t_{1} & \approx & \tilde{l}_{1}\,,\\
t_{2} & \approx & \left(2H\tilde{l}_{1}-2\dot{\phi}X\tilde{G}_{X}\right)\,,\\
t_{3} &\approx& 
-9\tilde{l}_{1}H^{2}+3\left(X\hat{\tilde{K}}_{X}+2X^{2}\hat{\tilde{K}}_{XX}\right)
\nonumber \\
& &{}+18H\dot{\phi}
\left(2X\tilde{G}_{X}+X^{2}\tilde{G}_{XX}\right)
-6(X G_{,\phi}+X^2 G_{,\phi X})\,,\\
t_{4}&\approx& \tilde{l}_{1}\,.
\end{eqnarray}

It is important to mention here that, in the action (\ref{eqxz}), both the coefficients of the terms 
$\alpha {\zeta}$ and ${\zeta}^2$ vanish by using the background equations of motion. 
Furthermore, in (\ref{eqxz}), the term quadratic in $\psi$ vanishes by making use of integrations by parts.
The equations of motion for $\psi$ and $\alpha$, derived from (\ref{eqxz}),
lead to the following two-fold constraint relations: 
\begin{eqnarray}\label{eqxc1}
\alpha &=& {\cal J} \dot{{\zeta}}\,,\\
\label{eqxc2}\frac{1}{a^2}\,\partial^2\psi &=& \frac{2t_3}{3t_2}\,\alpha
+3\dot{\zeta}-\frac{2t_1}{t_2}\frac1{a^2}
\partial^2{\zeta}\,,
\end{eqnarray}
where 
\begin{equation}
{\cal J} \equiv \frac{2t_1}{t_2}=\frac{2\tilde{l}_{1}}
{\left(2H\tilde{l}_{1}-2\dot{\phi}X\tilde{G}_{X}\right)}\,.
\end{equation}
Using expansion in terms of the slow-parameter defined in 
equation~(\ref{fifth}) gives 
\begin{equation}
{\cal J} =\frac{1}{H} \left[1+
\delta_{GX}+{\cal O} (\epsilon^2_{V}) \right]\,.
\label{L1expansion}
\end{equation}

Then substituting equation(\ref{eqxc1}) and equation(\ref{eqxc2})
into equation~(\ref{eqxz}) and integrating the term 
$\dot{\zeta}\partial^2{\zeta}$ by parts the second order  
action stated in equation(\ref{eqxz}) can be re-expressed as:
\be\label{ac2}
S^{\zeta}_{2}=\int dt d^{3}xa^{3}Y_{S}\left[\dot{\zeta}^{2}-\frac{c^{2}_{s}}{a^{2}}(\partial\zeta)^{2}\right],\ee
where 
\begin{eqnarray}\label{contr} Y_{S}&=&\frac{t_{1}\left(4t_{1}t_{3}+9t^{2}_{2}\right)}{3t^{2}_{2}}, \\
\label{contr1}c^{2}_{s}&=&\frac{3\left(2Ht_{2}t^{2}_{1}-t_{4}t^{2}_{2}-2t^{2}_{1}\dot{t}_{2}\right)}{t_{1}\left(4t_{1}t_{3}+9t^{2}_{2}\right)}.
\end{eqnarray}
 It is important to mention here that {\it ghosts} and {\it Laplacian} instabilities
can be avoided iff $c^{2}_{s}>0, Y_{S}>0$. Now using equation(\ref{eqxc1}) and equation(\ref{contr}) in equation(\ref{eqxc2}) we 
get:
\begin{eqnarray}\label{contr3} \psi &=& -{\cal J}\zeta+\partial^{-2}\left(\frac{a^{2}Y_{S}\dot{\zeta}}{t_{1}}\right).
\end{eqnarray}

 For future convenience, we have introduced a new parameter defined as:
\be\begin{array}{lll}\label{contr4}\displaystyle \epsilon_{s}=\frac{Y_{S}c^{2}_{s}}{\tilde{l}_{1}}=
\frac{\left(2Ht_{2}t^{2}_{1}-t_{4}t^{2}_{2}-2t^{2}_{1}\dot{t}_{2}\right)}{t^{2}_{2}\tilde{l}_{1}}=\epsilon_{V}+\delta_{GX}+{\cal O}(\epsilon^{2}_{V})
\end{array}\ee
where we use equation~(\ref{first}) and equation~(\ref{fifth}) to express it in terms of slow-roll parameters.
Now varying the action stated in equation(\ref{ac2}) and expressing the solution at the linear level in terms of Fourier modes,
we arrive at the {\it Mukhanov Sasaki Equation} for Galileon scalar mode. 
\be\label{ghy}
v^{''}_{\vec{k}}+\left(c^{2}_{s}k^{2}-\frac{z^{''}}{z}\right)v_{\vec{k}}=0,
\ee
where $c^2_{s}$ takes into account the nontrivial modification due to Galileon.
Similarly
for tensor modes, equation(\ref{ac2}) can be recast as:
\be\label{ac3}
S^{h}_{2}=\int dt d^{3}xa^{3}Y_{T}\left[\dot{h}^{2}_{ij}
-\frac{c^{2}_{s}}{a^{2}}(\partial h_{ij})^{2}\right],\ee
where 
\begin{eqnarray}\label{yui} Y_{T}&=&\frac{t_{1}}{4}=\frac{\tilde{l}_{1}}{4},\\ \label{yui8} c^{2}_{T}&=&\frac{t_{4}}{t_{1}}=1+{\cal O}(\epsilon^{2}).\end{eqnarray}
For tensor modes we use the normalization condition $e^{\lambda}_{ij}e^{\lambda^{'}}_{ij}=2\delta^{\lambda\lambda^{'}}$ and traceless condition $e_{ii}=0$ 
for polarization tensor.
Following the same prescription we can establish equation(\ref{ghy}) for tensor modes provided $c_{s}$
is replaced by $c_{T}$. 
The {\it Bunch-Davies} mode function turns out to be (Throughout the paper we
have used {\it DS} for {\it de-Sitter} results and {\it BDS} for {\it beyond de-Sitter} results.)
\be\label{ght}
u_{\zeta}(\eta,k)=
\left\{
	\begin{array}{ll}
                    \displaystyle\frac{iH\exp(-ikc_{s}\eta)}{2\sqrt{Y_{S}}(c_{s}k)^{\frac{3}{2}}}\left(1+ikc_{s}\eta\right)& \mbox{~~~~ \it:{\cal DS}}  \\
         \displaystyle  \frac{\sqrt{-k\eta c_{s}}}{a\sqrt{2Y_{S}}}{\cal H}^{(1)}_{\nu_{s}}(-k\eta c_{s})& \mbox{~~~~~\it:{\cal BDS }} .
          \end{array}
\right.
\ee
where $\nu_{s}=\left(\frac{3-\epsilon_{V}-2s^{S}_{V}+\delta_{V}}{2(1-
\epsilon_{V}-s^{S}_{V})}\right)$ and in the {\it super-Hubble}
 limit we have:
\begin{eqnarray}{\cal H}^{(1)}_{\nu_{s}}&\rightarrow&\frac{\left(-kc_{s}\eta\right)^{-\nu_{s}}
\exp\left(i[\nu_{s}-\frac{1}{2}]\frac{\pi}{2}\right)2^{\nu_{s}
-\frac{3}{2}}}{\sqrt{2c_{s}k}}\left(\frac{\Gamma(\nu_{s})}{\Gamma(\frac{3}{2})}\right)\end{eqnarray}.
 Further we have introduced five new parameters during the analysis of primordial 
quantum fluctuation defined as:
\bea \label{first1} s^{S}_{V}
:&=&\frac{\dot{c_{s}}}{Hc_{s}}=\frac{4\sqrt{V^{'}(\phi)}M^{2}_{PL}}{\sqrt{{\cal G}(\phi)}}\frac{d}{d\phi}\left(\ln c_{s}\right),
\\
\label{second2}s^{T}_{V} :&=&
\frac{\dot{c_{T}}}{Hc_{T}}=\frac{M_{PL}\sqrt{\tilde{g}_{1}V^{'}(\phi)}}{2\sqrt{e_{3}(\phi)V(\phi)}}\frac{d}{d\phi}\left(\ln\left[1+{\cal O}(\epsilon^2_{V})\right]\right),
\\
\label{third3} \eta_{s} :&=&
\frac{\dot{\epsilon_{s}}}{H\epsilon_{s}}=\frac{4\sqrt{V^{'}(\phi)}M^{2}_{PL}}{\sqrt{{\cal G}(\phi)}}\frac{\left[\epsilon^{'}_{V}\left(1+{\cal O}(\epsilon_{V})\right)+\delta_{GX\phi}
\right]}{\left[\epsilon_{V}+\delta_{GX}+{\cal O}(\epsilon^2_{V})\right]},
\\
\label{fourth4} \delta_{V} :&=&
\frac{\dot{Y_s}}{HY_{s}}=\frac{4\sqrt{V^{'}(\phi)}M^{2}_{PL}}{\sqrt{{\cal G}(\phi)}}\frac{d}{d\phi}\left(\ln Y_s\right),
\\
\label{fifth} \delta_{GX} :&=& \frac{\dot{\phi}X \tilde{G}_{X}}{\tilde{l}_{1}}.
\eea

Now using eqn(\ref{ght}) the {\it two-point} correlation function for scalar modes can be expressed as:
\be\label{tpt}
\langle0|\zeta(\vec{k})\zeta(\vec{k^{'}})|0\rangle=\frac{2\pi^{2}}{k^{3}}(2\pi)^{3}{\cal P}_{\zeta}(k)\delta^{3}(\vec{k}+\vec{k^{'}})
=(2\pi)^{3}|u_{\zeta}(\eta,k)|^{2}\delta^{3}(\vec{k}+\vec{k^{'}}),\ee
where the {\it dimensionless Power spectrum} for scalar modes ${\cal P}_{\zeta}(k)$ at the horizon crossing turns out to be:
\be\label{ght1}
{\cal P}_{\zeta}(k_{\star})=\frac{k^{3}_{\star}}{2\pi^{2}}|u_{\zeta}(k_{\star})|^{2}=
\left\{
	\begin{array}{ll}
                    \displaystyle\left(\frac{\sqrt{V(\phi)}}{8\pi^{2}c_{s}\epsilon_{s}\tilde{l}_{1}\sqrt{\tilde{g}_{1}}M_{PL}}\right)_{\star}
& \mbox{~ \it:{\cal DS}}  \\
         \displaystyle\left(2^{2\nu_{s}-3}\left|\frac{\Gamma(\nu_{s})}{\Gamma(\frac{3}{2})}\right|^{2}\frac{\left(1-\epsilon_{V}-s^{S}_{V}\right)^{2}\sqrt{V(\phi)}}{8\pi^{2}Y_{S}
c^{3}_{s}\sqrt{\tilde{g}_{1}}M_{PL}}\right)_{\star}& \mbox{~~\it:{\cal BDS }} .
          \end{array}
\right.
\ee
 $\star$ corresponds to the horizon crossing.
Similarly using the tensor version of eqn(\ref{ght}) the {\it two-point} correlation function for tensor modes can be expressed as:
\be\label{tptz}
\langle0|h_{ij}(\vec{k})h_{ij}(\vec{k^{'}})|0\rangle=\frac{2\pi^{2}}{k^{3}}(2\pi)^{3}{\cal P}_{T}(k)\delta^{3}(\vec{k}+\vec{k^{'}})
=(2\pi)^{3}|u_{\zeta}(\eta,k)|^{2}\delta^{3}(\vec{k}+\vec{k^{'}}),\ee
where ${\cal P}_{T}(k)=[{\cal P}_{T}(k)]_{ij;ij}$ and the corresponding {\it dimensionless Power spectrum} for tensor modes reads:
\be\label{ght2}
{\cal P}_{T}(k_{\star})=\frac{k^{3}_{\star}}{2\pi^{2}}|u_{h}(k_{\star})|^{2}\left(\sum_{\lambda = +,\times}e^{\lambda}_{ij}e^{\lambda}_{ij}\right)=
\left\{
	\begin{array}{ll}
                    \displaystyle\left(\frac{\sqrt{V(\phi)}}{2\pi^{2}c_{T}\epsilon_{T}\tilde{l}_{1}\sqrt{\tilde{g}_{1}}M_{PL}}\right)_{\star}
& \mbox{~ \it:{\cal DS}}  \\
         \displaystyle \left(2^{2\nu_{T}-3}\left|\frac{\Gamma(\nu_{T})}{\Gamma(\frac{3}{2})}\right|^{2}\frac{\left(1-\epsilon_{V}-s^{T}_{V}\right)^{2}\sqrt{V(\phi)}}{2\pi^{2}Y_{T}
c^{3}_{T}\sqrt{\tilde{g}_{1}}M_{PL}}\right)_{\star}& \mbox{~~\it:{\cal BDS }} .
          \end{array}
\right.
\ee
Consequently the ratio of tensor to scalar power spectrum can be expressed as:
\be\label{ght3}
{r}(k_{\star})=\frac{{\cal P}_{T}(k_{\star})}{{\cal P}_{\zeta}(k_{\star})}=\frac{|u_{h}(k_{\star})|^{2}
\left(\sum_{\lambda = +,\times}e^{\lambda}_{ij}e^{\lambda}_{ij}\right)}{|u_{\zeta}(k_{\star})|^{2}}=
\left\{
	\begin{array}{ll}
                    \displaystyle \left(16\epsilon_{s}c_{s}\left[1-\frac{3}{2}{\cal O}(\epsilon^{2}_{T})\right]\right)_{\star}
& \mbox{~ \it:{\cal DS}}  \\
         \displaystyle \left(16. 2^{2(\nu_{T}-\nu_{s})}\left|\frac{\Gamma(\nu_{T})}{\Gamma(\nu_{s})}\right|^{2}\left(\frac{1-\epsilon_{V}-s^{T}_{V}}
{1-\epsilon_{V}-s^{S}_{V}}\right)^{2}c_{s}\epsilon_{s}\left[1-\frac{3}{2}{\cal O}(\epsilon^{2}_{T})\right]\right)_{\star}& \mbox{~~\it:{\cal BDS }} .
          \end{array}
\right.
\ee
Further, the scale dependence of the perturbations, described by
the scalar and tensor spectral indices, as follows:
\be\label{ght3}
{n}_{\zeta}-1=\left(\frac{d\ln{\cal P}_{\zeta}}{d\ln k}\right)_{\star}=
\left\{
	\begin{array}{ll}
                    \displaystyle \left(-2\epsilon_{V}-\eta_{s}-s^{S}_{V}\right)_{\star}
=\left(-2\epsilon_{s}-\eta_{s}-s^{S}_{V}+2\delta_{GX}+2{\cal O}(\epsilon^{2}_{s})\right)_{\star}& \mbox{ \it:{\cal DS}}  \\
         \displaystyle  \left(3-2\nu_{s}\right)=-\left(\frac{2\epsilon_{V}+s^{S}_{V}+\delta_{V}}{1-\epsilon_{V}-s^{S}_{V}}\right)_{\star}& \mbox{\it:{\cal BDS }} .
          \end{array}
\right.
\ee
\be\label{ght4}
{n}_{T}=\left(\frac{d\ln{\cal P}_{T}}{d\ln k}\right)_{\star}=
\left\{
	\begin{array}{ll}
                    \displaystyle -2\epsilon_{V}|_{\star}
=\left(-2\epsilon_{s}+2\delta_{GX}+2{\cal O}(\epsilon^{2}_{s})\right)_{\star}& \mbox{ \it:{\cal DS}}  \\
         \displaystyle  \left(3-2\nu_{T}\right)=-\left(\frac{s^{T}_{V}}{1-\epsilon_{V}-s^{T}_{V}}\right)_{\star}& \mbox{\it:{\cal BDS }} .
          \end{array}
\right.
\ee
Consistently, the consistency relation is also modified to:
\be\label{ght5}
{r}=
\left\{
	\begin{array}{ll}
                    \displaystyle\left(8c_{s}\left(2\epsilon_{V}+2\delta_{GX}
+2{\cal O}(\epsilon^{2}_{V})\right)\left[1-\frac{3}{2}{\cal O}(\epsilon^{2}_{T})\right]\right)_{\star}
=-\left(8c_{s}\left(n_{T}-2\delta_{GX}
-2{\cal O}\left(\frac{n^{2}_{T}}{4}\right)\right)\left[1-\frac{3}{2}{\cal O}(\epsilon^{2}_{T})\right]\right)_{\star}
& \mbox{~ \it:{\cal DS}}  \\
         \displaystyle\left(8. 2^{2(\nu_{T}-\nu_{s})}\left|\frac{\Gamma(\nu_{T})}{\Gamma(\nu_{s})}\right|^{2}\left(\frac{1-\epsilon_{V}-s^{T}_{V}}
{1-\epsilon_{V}-s^{S}_{V}}\right)^{2}c_{s}\left(2\epsilon_{V}+2\delta_{GX}+2{\cal O}\left(\epsilon^{2}_{V}\right)\right)
\left[1-\frac{3}{2}{\cal O}(\epsilon^{2}_{T})\right]\right)_{\star}\\
\displaystyle =\left(8. 2^{(n_{\zeta}-n_{T})}\left|\frac{\Gamma(\frac{3-n_{T}}{2})}{\Gamma(\frac{3-n_{\zeta}}{2})}\right|^{2}\left(\frac{\frac{s^{T}_{V}}{n_{T}}}
{s^{T}_{V}\left(1-\frac{1}{n_{T}}\right)-s^{S}_{V}}\right)^{2}c_{s}\left(2\left[1+s^{T}_{V}\left(\frac{1}{n_{T}}-1\right)\right]
\right.\right.\\ \left.\left.\displaystyle ~~~~~~~~~~~~~~~~~~~~~~~~~~~~~~~~~~~~~~~~~~~~~~~~~~~~~~~~~~~~~~~~~~~~~~~~~~~~~~~~+2\delta_{GX}+2{\cal O}(\epsilon^{2}_{V})\right)
\left[1-\frac{3}{2}{\cal O}(\epsilon^{2}_{T})\right]\right)_{\star} & \mbox{~~\it:{\cal BDS }} .
          \end{array}
\right.
\ee

The expressions for the
running of the scalar and tensor spectral index in this specific 
model with respect to the logarithmic pivot scale at the horizon crossing are given by:

\be\label{ght6}
{\alpha}_{\zeta}=\left(\frac{d{n}_{\zeta}}{d\ln k}\right)_{\star}=
\left\{
	\begin{array}{ll}
                    \displaystyle \left\{\frac{4\sqrt{V^{'}(\phi)}M^{2}_{PL}}{\sqrt{{\cal G}(\phi)}}\left(-2\epsilon^{~'}_{V}-\eta^{'}_{s}-s^{S~'}_{V}\right)\right\}_{\star}& \mbox{ \it:{\cal DS}}  \\
         \displaystyle  \frac{4\sqrt{V^{'}(\phi)}M^{2}_{PL}}{\sqrt{{\cal G}(\phi)}\left(1-\epsilon_{V}-s^{S}_{V}\right)^{2}}
\left[\underbrace{s^{S~'}_{V}\epsilon^{~'}_{V}}+\underbrace{\delta^{~'}_{V}\epsilon^{~'}_{V}}
+\underbrace{s^{S~'}_{V}\delta^{~'}_{V}}+\left(2\epsilon^{~'}_{V}+s^{S~'}_{V}+\delta^{~'}_{V}\right)\right]& \mbox{\it:{\cal BDS }} .
          \end{array}
\right.
\ee
\be\label{ght7}
{\alpha}_{T}=\left(\frac{d{n}_{T}}{d\ln k}\right)_{\star}=
\left\{
	\begin{array}{ll}
                    \displaystyle -2\left\{\frac{4\sqrt{V^{'}(\phi)}M^{2}_{PL}}{\sqrt{{\cal G}(\phi)}}\epsilon^{~'}_{V}\right\}_{\star}& \mbox{ \it:{\cal DS}}  \\
         \displaystyle -\left[\frac{4\sqrt{V^{'}(\phi)}M^{2}_{PL}}{\sqrt{{\cal G}(\phi)}}
\left(\frac{s^{T~'}_{V}}{1-\epsilon_{V}-s^{T}_{V}}\right)+\frac{s_{T}\left(\epsilon^{~'}_{V}+s^{T~'}_{V}\right)}{\left(1-\epsilon_{V}-s^{T}_{V}\right)^{2}}\right]_{\star}& \mbox{\it:{\cal BDS }} .
          \end{array}
\right.
\ee
 
Here we have used a shorthand notation $\underbrace{ab}=a^{'}b-ab^{'}$ where $'=\frac{d}{d\phi}$.
We also use the operator identy $\frac{d}{d\ln k}:=\frac{4\sqrt{V^{'}(\phi)}M^{2}_{PL}}{\sqrt{{\cal G}(\phi)}}\frac{d}{d\phi}$ to compute all the inflationary observables.


 Figure(\ref{fig:subfigureExample1}(a)) represents the scale dependent power spectrum 
($\sqrt{{\cal P}_{\zeta}}$) with respect to the scalar spectral index($n_{\zeta}$). It directly 
shows that both the {\it DS} and {\it BDS} analysis
follow the same characteristics but the estimated windows for the 
observational parameters (${\cal P}_{\zeta},n_{\zeta}$) are slightly different,
but both of them are within the observational bound.
Figure(\ref{fig:subfigureExample1}(b)) shows the characteristic
differences between the behavior of {\it DS} and {\it BDS} scale dependent 
power spectrum with respect to the momentum scale ($k$). Here {\it DS} behavior
is quasi-statically flat, but {\it BDS} characteristics is rapidly increasing with 
respect to the scale. In figure(\ref{fig:subfigureExample1}(c)) we have
plotted the the scale dependent tensor to scalar ratio for
{\it DS} and {\it BDS} limit. Most significantly they show complementary 
characteristics with the scale and intersects at a point where both the analysis
will be equivalent. In the next section, we will estimate these parameters by confronting the results directly to WMAP7
results.

\section{\bf Parameter Estimation and Confrontation with WMAP7}
 
Using the parameter space for the model parameters ($C_{i},D_{i}$) 
we have estimated the window of the cosmological parameters from our model which confronts
 observational data well in $56<N<70$. In Table(\ref{tab21}) we have tabulated the relevant observational parameters 
estimated from our model for both {\it DS} and {\it BDS} limit.

\begin{table}[h]
\begin{tabular}{|c|c|c|c|c|c|c|c|c|c|}
\hline Scheme & ${\cal P}_{\zeta}$ & $r$ &$n_{\zeta}$& $\alpha_{\zeta}$
 \\  &$\times 10^{-9}$
 & & & $\times \left(-10^{-3}\right)$\\
 \hline
DS&~2.401~-~2.601~&~0.215~-~0.242~&~0.964~-~0.966~&~2.240~-~2.249~\\
\hline
BDS&~ 2.471~-~2.561~&~0.232~-~0.250~&~0.962~-~0.964~&~4.008~-~4.012~\\
\hline
\end{tabular}
\caption{Model Dependent Observational Parameters}\label{tab21}
\end{table}

\begin{figure}[h]
\centering
\subfigure[]{
    \includegraphics[width=5.6cm,height=5.6cm] {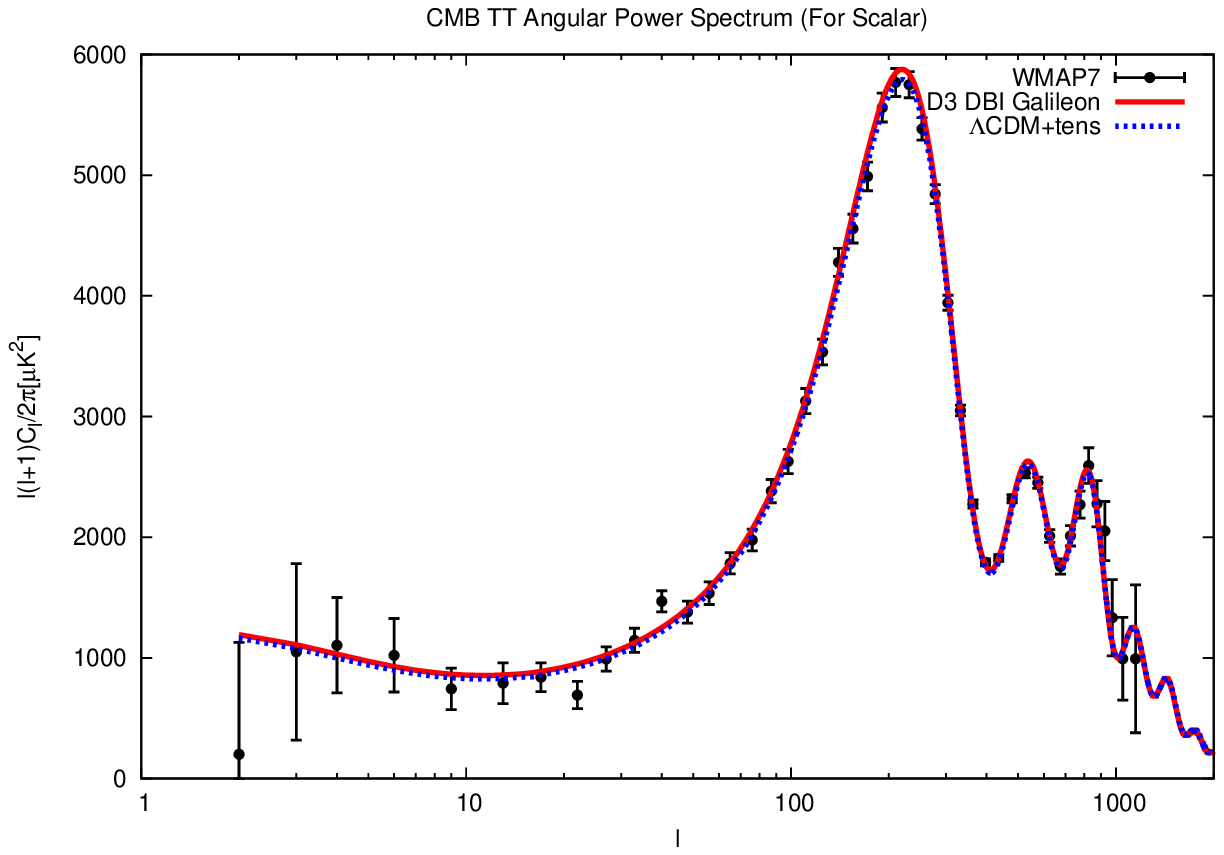}
    \label{fig:subfig5}
}
\subfigure[]{
    \includegraphics[width=5.6cm,height=5.6cm] {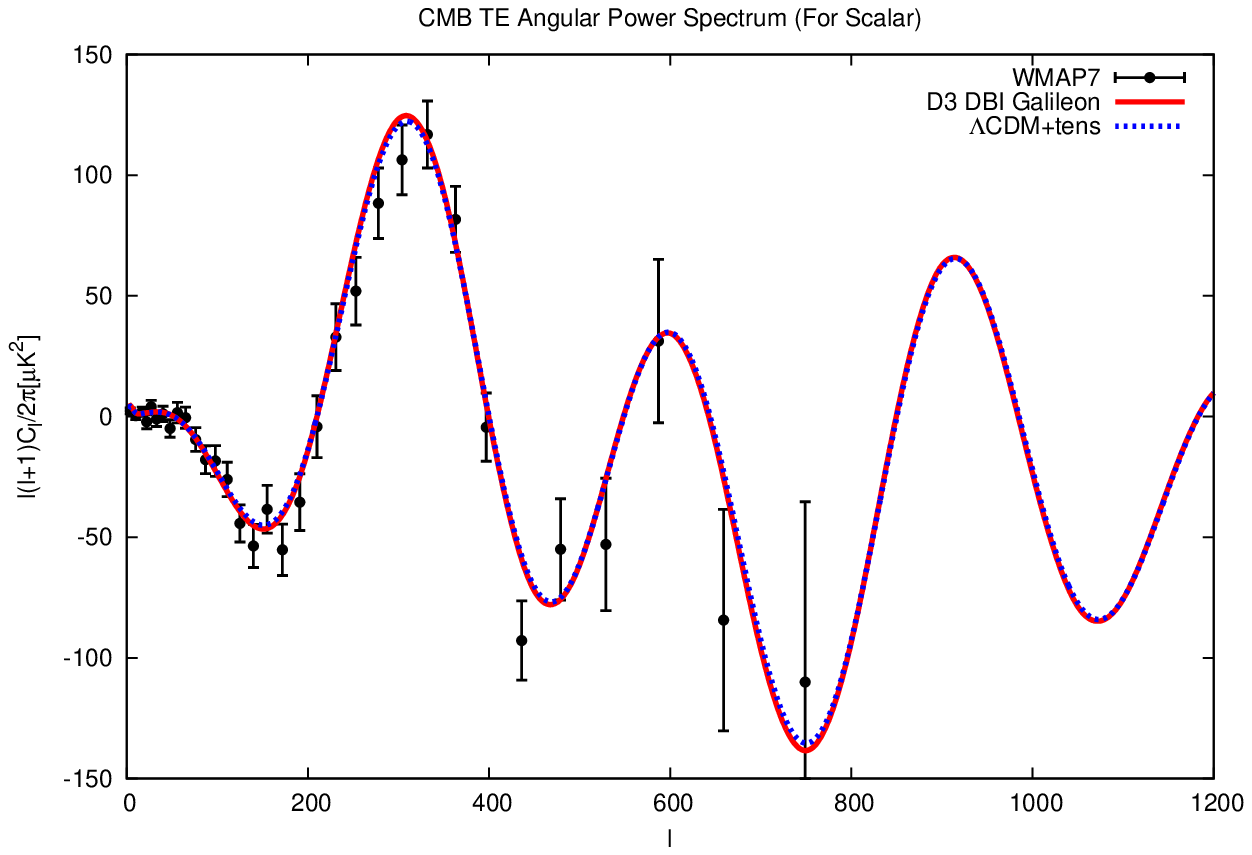}
    \label{fig:subfig6}
}
\subfigure[]{
    \includegraphics[width=5.6cm,height=5.6cm] {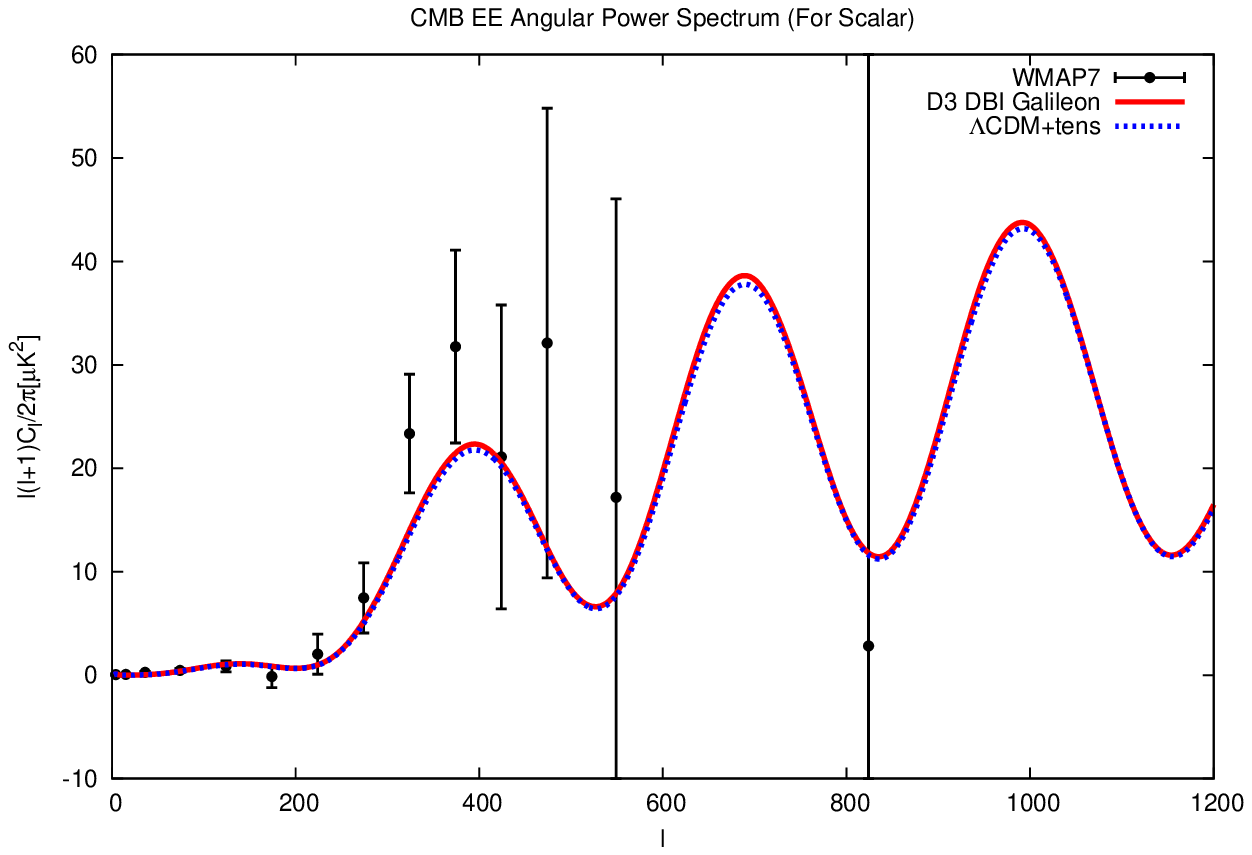}
    \label{fig:subfig7}
}
\subfigure[]{
    \includegraphics[width=5.6cm,height=5.6cm] {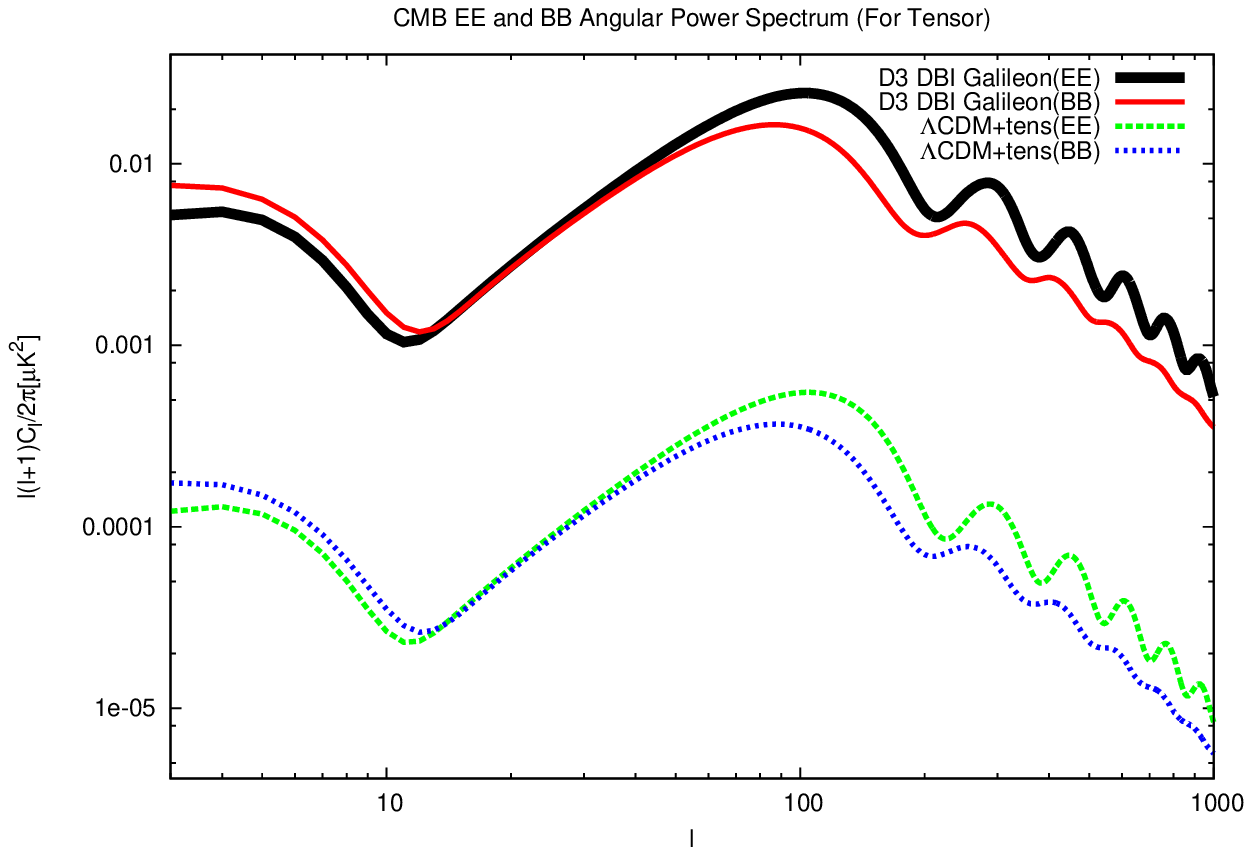}
    \label{fig:subfig8}
}
\subfigure[]{
    \includegraphics[width=5.6cm,height=5.6cm] {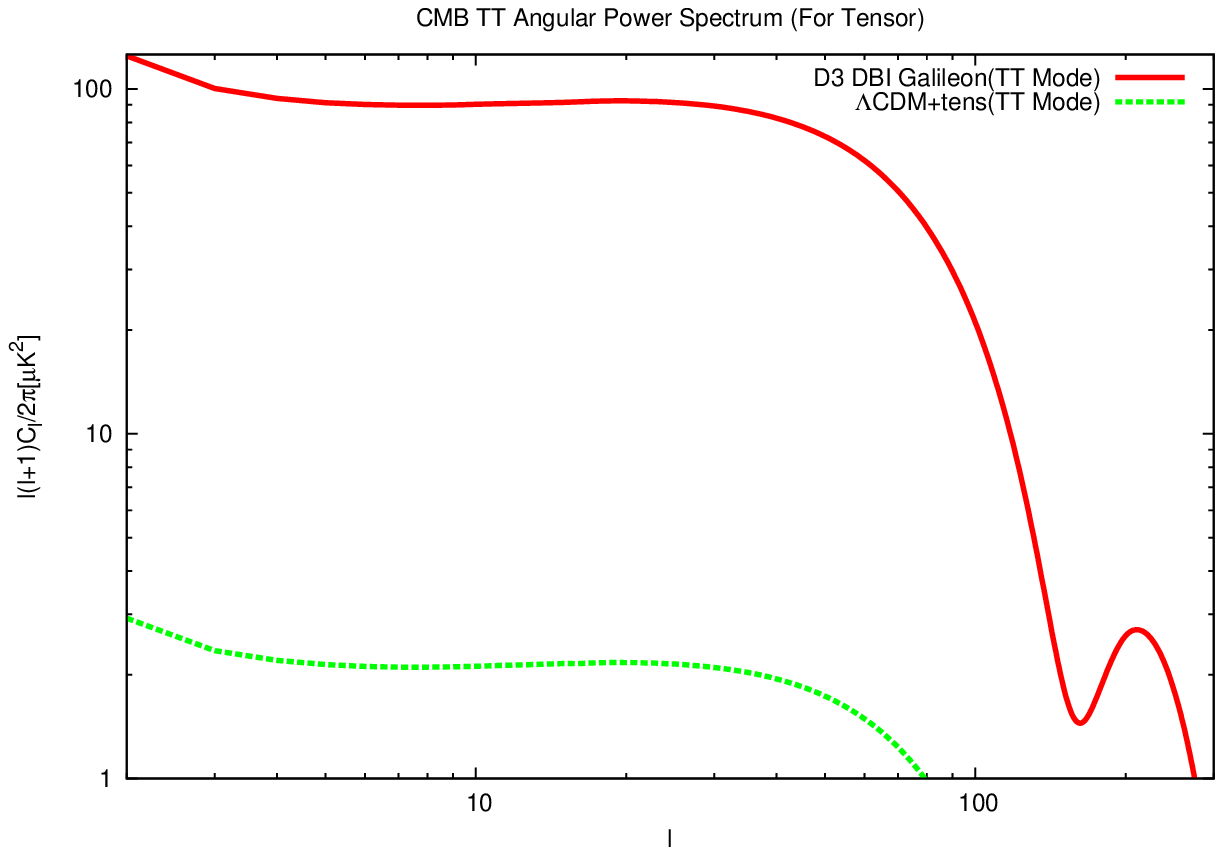}
    \label{fig:subfig9}
}
\subfigure[]{
    \includegraphics[width=5.6cm,height=5.6cm] {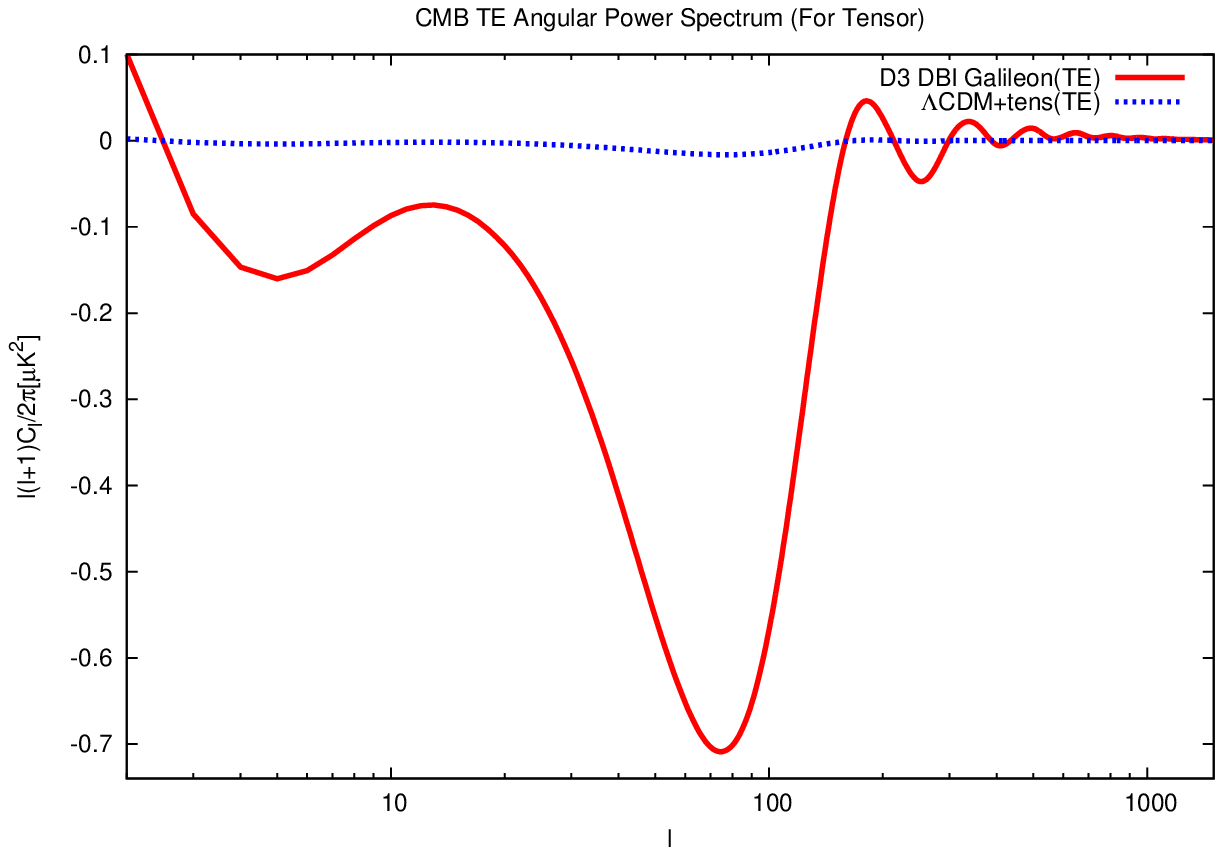}
    \label{fig:subfig15}
}
\caption{ Variation
 of the CMB {\subref{fig:subfig5}TT (scalar), \subref{fig:subfig6}TE (scalar), \subref{fig:subfig7}EE (scalar),
\subref{fig:subfig8}EE+BB (tensor), \subref{fig:subfig9}TT (tensor) and \subref{fig:subfig15}TE (tensor) angular power spectrum with multipoles ($l$).}
\label{fig:subfigureExample3}}
\end{figure}
\begin{figure}[h]
{\centerline{\includegraphics[width=10cm, height=8.1cm] {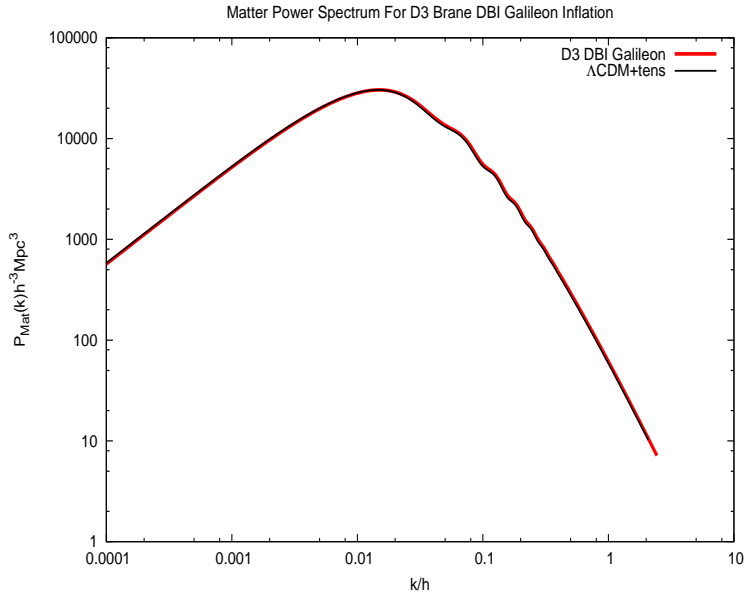}}}
\caption{Variation of matter power spectrum vs momentum scale k} \label{figVrwe4}
\end{figure}

\begin{table}[h]
\parbox{.46\linewidth}{
\centering
\begin{tabular}{|c|c|c|c|c|c|c|c|c|c|}
\hline $H_0$ & $\tau_{Reion}$ &$\Omega_b h^2$& $\Omega_c h^2
$& $T_{CMB}$
 \\
km/sec/MPc& & && K\\
 \hline
71.0&0.09&0.0226&0.1119&2.725\\
\hline
\end{tabular}
\caption{Input parameters in CAMB}\label{tab2}
}
\hfill
\parbox{.46\linewidth}{
\centering
\begin{tabular}{|c|c|c|c|c|c|c|c|c|c|}
\hline $t_0$ & $z_{Reion}$ &$\Omega_m$&$\Omega_{\Lambda}$&$\Omega_k$&$\eta_{Rec}$& $\eta_0$
 \\
Gyr& & && &Mpc & Mpc\\
 \hline
13.707&10.704&0.2670&0.7329&0.0&285.10&14345.1\\
\hline
\end{tabular}
\caption{Output parameters from CAMB}\label{tab3}
}
\end{table}
Further, we use the publicly available code CAMB \cite{camb}
to verify our results directly with observation. To
operate CAMB at the pivot scale $k_{0} = 0.002 Mpc^{-1}$
the values of the initial parameter space are taken for
 ̃lower bound of $C_{i}^{'}$s and $N = 70$. Additionally
WMAP7 years dataset for $\Lambda CDM$ background has
been used in CAMB to obtain CMB angular power spectrum. In Table(\ref{tab2}) we have given all the input parameters
for CAMB. Table(\ref{tab3}) shows the CAMB output, which is
in good agreement with WMAP7 \cite{wmap7} data. In
figure(\ref{fig:subfigureExample3})(a)-figure(\ref{fig:subfigureExample3})(c) we have plotted CAMB output of CMB TT, TE and EE
 angular power spectrum $C^{TT}_{l}, C^{TE}_{l}, C^{EE}_{l}$ for the best fit with WMAP7
 data for scalar mode, which explicitly show the
agreement of our model with WMAP7 dataset. The small scale modes have no
impact in the CMB anisotropy spectrum only the large scale modes
have little contribution and this is obvious from figure(\ref{fig:subfigureExample3})(d)-figure(\ref{fig:subfigureExample3})(f) where 
we have plotted the CAMB output of CMB angular power spectrum $C_l^{TT}$, $C_l^{TE}$, $C_l^{EE}$ and $C_l^{BB}$
 for best fit with WMAP7 data for the tensor mode. Hence in figure(\ref{figVrwe4})we
 have plotted the variation of matter power spectrum
with respect to the momentum scale which is in concordance with observational results.


\section{\bf Summary and Outlook}

In this article we have proposed a model of single field inflation in the context of DBI Galileon
cosmology in D3 brane. We have demonstrated the technical details of construction mechanism of an one-loop
4D inflationary potential via dimensional reduction starting from D4 brane in ${\cal N}$=2,${\cal D}$=5 SUGRA including 
the quadratic {\it Gauss-Bonnet} correction term that leads to an effective ${\cal N}$=1,${\cal D}$=4 SUGRA in the D3 brane, which is precisely the DBI Galileon in our framework.
 Hence 
we have studied inflation using the one loop effective potential
by estimating the observable parameters originated from primordial quantum fluctuation for scalar and tensor modes, in the de-Sitter and beyond de-sitter limit.
We have further confronted our results with WMAP7 \cite{wmap7} dataset by
using the cosmological code CAMB. 
The results are found in good agreement with WMAP7 dataset in {\it $\Lambda$CDM+tens} background.

An interesting open issue in this context is to study the primordial non-Gaussian features of DBI Galileon introduced in the present article.
 As has been pointed out recently \cite{tye1}-\cite{tye3} there is a tension between bispectrum ($f_{NL}$) and tensor-to-scalar ratio ($r$)
in DBI inflation, which is a generic sensitivity problem. It will be interesting to investigate whether our proposed framework of DBI Galileon can resolve this issue.
We are already in progress in this direction and  have obtained some interesting results. A detailed report
on this issue will be brought forth shortly.

 Other open issues in the context of
DBI Galileon cosmology are combined constraints on the primordial
 non-Gaussianity via preheating \cite{dhiraj}, reheating \cite{teru} and primordial black hole formation \cite{green},
 effect of the presence of one loop and two loop radiative corrections in the 
presence of all possible scalar and tensor mode fluctuations up to the fourth order correction in the action \cite{sanu},
 study of different shapes and comparative study between the tree, one and two loop level via rigorous analysis and
 finding out the most probable Dark Matter candidate in collider phenomenology.
It will be interesting to see how they are affected due to the presence of Galileon.


\section*{\bf Acknowledgments}

SC thanks Council of Scientific and
Industrial Research, India for financial support through Senior
Research Fellowship (Grant No. 09/093(0132)/2010).

\section*{\bf Appendix A}
\label{apa}


In this section we employ dimensional reduction technique to derive 
a ${\cal N}$=1, ${\cal D}$=4 SUGRA and the inflaton potential therefrom that results in DBI Galileon on the D3 brane. For convenience we deal with different contributions to the action
(\ref{totac}) separately.\\ 
\\
\underline{\bf The Einstein-Hilbert Action}:-\\
\\
After integrating out the contribution from the five dimension, the  Einstein Hilbert action in four dimension can be written as
\be\begin{array}{llllll}\label{ghu}\displaystyle S^{(4)}_{EH}=\frac{1}{2}\int d^{4}x\sqrt{-g_{(4)}}\int^{+\pi R}_{-\pi R}dy \beta M^{3}_{5}R\exp(3A(y))
\left[R_{(4)}-\frac{12}{\beta^{2}R^{2}}\left(\frac{dA(y)}{dy}\right)^{2} 
-\frac{8}{\beta^{2}R^{2}}\left(\frac{d^{2}A(y)}{dy^{2}}\right)-2\Lambda_{5}\exp(2A(y))\right]\\
~~~~~~=\displaystyle\frac{1}{2{\kappa}^{2}_{4}}\int d^{4}x\sqrt{-g_{(4)}}\left[R_{(4)}-\frac{3M^{3}_{5}\beta b^{6}_{0}}{M^{2}_{PL}R^{5}}
{\cal I}(1)\right],\end{array}\ee
where the explicit expression for ${\cal I}(1)$ is mentioned in Appendix C. In this context $R_{(4)}$ is the 4D Ricci scalar. It is important to mention here that
the 5D Planck mass ($M_{5}$) and 4D Planck mass ($M_{PL}$) are related through 
\be\begin{array}{lllll}\label{mass}
   \displaystyle M^{2}_{PL}=M^{3}_{(5)}\beta R\int^{+\pi R}_{-\pi R}dy \exp(3A(y))\\
~~~~~~~=\frac{M^{3}_{(5)}b^{3}_{0}}{3R^{2}T^{3/2}_{(4)}}\left[\exp(\beta \pi R)
\left\{\frac{3\sqrt{T_{(4)}}}{\sqrt{\exp(\beta\pi R)+T_{(4)}\exp(-\beta\pi R)}}
-\sqrt{\frac{\exp(2\pi\beta R)+T_{(4)}}{\exp(\beta\pi R)+T_{(4)}\exp(-\beta\pi R)}}
~_2F_1\left[\frac{1}{2};\frac{3}{7};\frac{7}{4};-\frac{\exp(2\beta\pi R)}{T_{(4)}}\right]\right\}
\right.\\ \left.~~~~~~~~~~~~~~~~~~~~~~-\exp(-\beta\pi R)\left\{\frac{3\sqrt{T_{(4)}}}{\sqrt{\exp(-\beta\pi R)+T_{(4)}\exp(\beta\pi R)}}
-\sqrt{\frac{\exp(-2\beta\pi R)+T_{(4)}}{\exp(-\beta\pi R)+T_{(4)}\exp(\beta\pi R)}}
~_2F_1\left[\frac{1}{2};\frac{3}{7};\frac{7}{4};-\frac{\exp(-2\beta\pi R)}{T_{(4)}}\right]\right\}\right].
   \end{array}
\ee 


\underline{\bf The Gauss-Bonnet Action}:-\\

The 5D and 4D {\it Riemann tensor}, {\it Ricci tensor} and {\it Ricci scalar} are related through the following expressions:
\be\begin{array}{lllll}\label{r1}
  \displaystyle R^{(5)}_{\alpha\beta\gamma\delta}=R^{(4)}_{\alpha\beta\gamma\delta}
+\frac{\exp(2A(y))}{R^{2}\beta^{2}}\left(\frac{dA(y)}{dy}\right)^{2}\left[g^{(4)}_{\gamma\beta}g^{(4)}_{\alpha\delta}
-g^{(4)}_{\alpha\gamma}g^{(4)}_{\delta\beta}\right]
   \end{array}
\ee
\be\begin{array}{llll}\label{r2}
    \displaystyle R^{(5)}_{\alpha\beta}=R^{(4)}_{\alpha\beta}
-\frac{3g^{(4)}_{\alpha\beta}}{R^{2}\beta^{2}}\left(\frac{dA(y)}{dy}\right)^{2}
   \end{array}
\ee
\be\begin{array}{llll}\label{r5}
    \displaystyle R_{(5)}=\exp(2A(y))\left[R_{(4)}-\frac{12}{\beta^{2}R^{2}}\left(\frac{dA(y)}{dy}\right)^{2}
-\frac{8}{\beta^{2}R^{2}}\left(\frac{d^{2}A(y)}{dy^{2}}\right)-2\Lambda_{5}\exp(2A(y))\right]
   \end{array}
\ee

Using eqn(\ref{r1})-eqn(\ref{r5}) in eqn(\ref{5gb}) we get 
\be\begin{array}{lllll}\label{gbe4}
     \displaystyle S^{(4)}_{GB}
=\frac{\alpha_{(4)}}{2\kappa^{2}_{(4)}}\int d^{4}x\sqrt{-g_{(4)}}
\left[\left({\cal C}(1)R^{\alpha\beta\gamma\delta(4)}R^{(4)}_{\alpha\beta\gamma\delta}
-4{\cal I}(2)R^{\alpha\beta(4)}R^{(4)}_{\alpha\beta}+{\cal A}(6)R^{2}_{(4)}\right)\right.\\ \left.
\displaystyle ~~~~~~~~~~~~~~~~~~~~~~~~~~~~+\frac{2{\cal C}(2)}{R^{2}\beta^{2}}R^{(4)}_{\alpha\beta\gamma\delta}\left(g^{\gamma\beta(4)}g^{\delta\alpha(4)}
-g^{\gamma\alpha(4)}g^{\delta\beta(4)}\right)
+\frac{{\cal G}(1)}{R^{4}\beta^{4}}+\frac{{\cal G}(2)}{R^{2}\beta^{2}}R_{(4)}\right]
   \end{array}
\ee
where ${\cal G}(1)=24{\cal C}(4)-144{\cal I}(4)-64{\cal A}(5)+144{\cal A}(7)+64{\cal A}(8)+192{\cal A}(11)$ and
${\cal G}(2)=24{\cal I}(2)-24{\cal A}(9)-16{\cal A}(10)$. The scaling relationship between 4D and 5D {\it Gauss-Bonnet}
coupling constant is $\alpha_{(4)}=\frac{\kappa^{2}_{(4)}\beta R}{\kappa^{2}_{(5)}}\alpha_{(5)}$ where $\kappa_{(4)}$ and $\kappa_{(5)}$ are 
gravitational couplings in 4D and 5D respectively.
 Explicit form of each of the constants appearing in eqn(\ref{gbe4}) are mentioned in the 
Appendix C.
\\
\\
\underline{\bf The D3 Brane Action}:-\\

To reduce the D4 brane action we employ the method of separation of variable 
$\Phi(X^{A})=\Phi(x^{\mu},y)=\phi(x^{\mu})\chi(y)$ where $\chi(y)=\exp(\frac{2\pi i y}{R})$. Consequently the  D3 
brane action turns out to be 
\be\label{br1}
   S^{(4)}_{D3~Brane}= \int d^{4}x \sqrt{-g^{(4)}}
\left[\tilde{K}(\phi,\tilde{X})-\tilde{G}(\phi,\tilde{X})\Box^{(4)}\phi\right],\ee  
where $\tilde{K}(\phi,X)=\left\{-\frac{\tilde{D}}{\tilde{f}(\phi)}\left[\sqrt{1-2Q\tilde{X}\tilde{f}}-Q_{1}\right]
-\tilde{C}_{5}\tilde{G}(\phi,\tilde{X})-\bar{Q}_{2}\tilde{D}V^{(4)}_{brane}\right\}$, 
$\tilde{G}(\phi,\tilde{X})=\left(\frac{\tilde{g}(\phi)k_{1}\tilde{C}_{4}}{2(1-2\tilde{f}(\phi)\tilde{X}k_{2}))}\right)$,
$\tilde{g}(\phi)=\tilde{g}_{0}+\tilde{g}_{2}\phi^{2}$,
$\tilde{D}=\frac{D}{2\kappa^{2}_{(4)}}$, $\tilde{C}_{4}=\frac{C_{4}}{2\kappa^{2}_{4}}$, 
$\tilde{C}_{5}=\frac{\bar{C}_{5}\beta^{2} R^{2}}{2\kappa^{2}_{4}}$, $\bar{Q}_{2}\tilde{D}=\beta R$. The effective {\it Klebanov Strassler}
and {\it Coulomb } frame function on the D3 brane are hereby expressed as 
$\tilde{f}(\phi)\simeq\frac{1}{(\tilde{f}_{0}+\tilde{f}_{2}\phi^{2}+\tilde{f}_{4}\phi^{4})}$ and 
$\nu^{(4)}(\phi)=\tilde{\nu}_{0}+\frac{\tilde{\nu}_{4}}{\phi^{4}}$ with $\tilde{\nu}_{0}=\nu_{0}{\cal A}(1)$
, $\tilde{\nu}_{4}=\nu_{4}{\cal A}(12)$. The outcome of dimensional reduction is reflected through the constants mentioned in the Appendix C.
The scaled D3 brane potential turns out to be
\be\label{scaled} \tilde{V}^{(4)}_{brane}=\bar{Q}_{2}\tilde{D}V^{(4)}_{brane}=
T^{}_{(3)}\nu^{(4)}(\phi)+\frac{\beta R{\cal I}(2)}{\tilde{f}(\phi)}\ee
where the  D3 brane tension ($T_{(3)}$) can be expressed in terms of the D4 brane tension ($T_{(4)}$), compactification radius ($R$) and
the slope parameter ($\beta$) as  $T^{}_{(3)}=\beta R T_{(4)}$.
\\
\\
 \underline{\bf The ${\cal N}$=1, ${\cal D}$=4 Supergravity Action}:-\\

Further, imposing $Z_{2}$ symmetry to $\phi$ via
$\Phi(0)=\Phi(\pi
R)=0$ and compactifying around a circle $(S^1)$
$\partial_{5}\Phi=\sqrt{V^{(5)}_{bulk}(G)}\left(1-\frac{1}{2\pi
R}\right)$ we get,
\be \label{tout}
S^{(5)}_{Bulk~Sugra}=\frac{1}{2}\int d^{4}x\int^{+\pi R}_{-\pi
R}dy\sqrt{-g^{(5)}}\left[e_{(4)}e^{5}_{\dot{5}}\left\{g^{\alpha\beta}G_{m}^{n}(\partial_{\alpha}\phi^{m})^{\dagger}(\partial_{\beta}\phi_{n})
-g^{55}\frac{V^{(5)}_{bulk}(G)}{4\pi^{2}R^{2}}\right\}\right].\ee
Now using the above mentioned ansatz for method of separation of variable we get
\be\begin{array}{lllll}\label{ast8}\displaystyle S^{(4)}_{Sugra}=
\frac{1}{2\kappa^{2}_{(4)}}\int d^{4}x \sqrt{-g^{(4)}}\left[M(T,T^{\dag})
{\cal J}_{\nu}^{\mu}(\phi,\phi^{\dag})g^{\alpha\beta (4)}(\partial_{\alpha}\phi_{\mu})^{\dag}(\partial_{\beta}\phi^{\nu})
-Z(T,T^{\dag})V^{(4)}_{F}(\phi)
\right].\end{array}\ee
where we define

\be\begin{array}{lllll}\label{talx}\displaystyle {\cal J}_{\nu}^{\mu}(\phi,\phi^{\dag})=\int^{+\pi R}_{-\pi R}dy \exp(-A(y))
\left(\frac{\partial^{2}{\cal K}(\phi\exp(\frac{2\pi i y}{R}), 
\phi^{\dag}\exp(-\frac{2\pi i y}{R}))}{\partial\phi^{\dag}_{\mu}\partial\phi^{\nu}}\right), \\ \displaystyle
M(T,T^{\dag})=\frac{\sqrt{2}\beta R^{2}}{(T+T^{\dag})}, Z(T,T^{\dag})=\frac{1}{8\pi^{2}R^{2}\beta|T+T^{\dag}|^{2}}.\end{array}\ee 
Here we have used the ansatz ${\cal W}(\phi,\phi^{\dag},T,T^{\dag})=\frac{{\cal W}_{1}(\phi,\phi^{\dag})|T+T^{\dag}|^{2}}{4}$ 
for superpotential.
Hence, the  ${\cal N}=1,{\cal  D}=4$
supergravity F-term potential turns out to be
\be\begin{array}{llllllll}\label{vf}
\displaystyle V^{(4)}_{F}=\int^{+\pi R}_{-\pi R}dy\exp(-A(y))\exp\left(\frac{{\cal K}(\phi\exp(\frac{2\pi i y}{R}),
\phi^{\dag}\exp(-\frac{2\pi i y}{R}))}{M^{2}}\right)\\~~~~~~~~~~~~~~~~~~~~~~~~~~\displaystyle \left[\left(\frac{\partial{\cal W}_{1}}{\partial
\phi_{\alpha}}+\left(\frac{\partial {\cal K}(\phi\exp(\frac{2\pi i y}{R}),
\phi^{\dag}\exp(-\frac{2\pi i y}{R}))}{\partial
\phi_{\alpha}}\right)\frac{{\cal W}_{1}}{M^{2}}\right)^{\dag}\left(\frac{\partial^{2}{\cal K}(\phi\exp(\frac{2\pi i y}{R}),
\phi^{\dag}\exp(-\frac{2\pi i y}{R}))}{\partial\phi^{\alpha}\partial\phi^{\dagger}_{\beta}}\right)^{-1}\right.\\ \left.
~~~~~~~~~~~~~~~~~~~~~~~~~~~~~~~~~~~~~~~~~~~~~~~~~~~~~~~~\displaystyle\left(\frac{\partial {\cal W}_{1}}{\partial
\phi^{\beta}}+\left(\frac{\partial {\cal K}(\phi\exp(\frac{2\pi i y}{R}),
\phi^{\dag}\exp(-\frac{2\pi i y}{R}))}{\partial
\phi^{\beta}}\right)\frac{{\cal W}_{1}}{M^{2}}\right)-3\frac{|{\cal W}_{1}|^{2}}{M^{2}}\right].\end{array}\ee
Now using the ansatz for the {\it K$\ddot{a}$hler } potential ${\cal K}(\phi\exp(\frac{2\pi i y}{R}),
\phi^{\dag}\exp(-\frac{2\pi i y}{R}))
={\cal K}_{1}(\phi,\phi^{\dag}){\cal K}_{2}(\exp(\frac{2\pi i y}{R}),\exp(-\frac{2\pi i y}{R}))$
with ${\cal K}_{1}(\phi,\phi^{\dag})={\cal K}_{1}^{\alpha\beta}\phi_{\alpha}\phi^{\dag}_{\beta}$
and ${\cal K}_{2}(\exp(\frac{2\pi i y}{R}),\exp(-\frac{2\pi i y}{R}))=1$ eqn(\ref{vf}) reduces to the following form:
\be\begin{array}{lllllll}\label{exact}
 \displaystyle V^{(4)}_{F}= {\cal A}(13)\exp\left(\frac{{\cal K}_{1}^{\alpha\beta}\phi_{\alpha}\phi^{\dag}_{\beta}}{M^{2}}\right)
\left[\left(\frac{\partial {\cal W}_{1}}{\partial
\phi_{\alpha}}+{\cal K}_{1}^{\alpha\beta}\phi^{\dag}_{\beta}\frac{{\cal W}_{1}}{M^{2}}\right)^{\dag}
 {\cal K}_{1\alpha}^{\nu}\left(\frac{\partial {\cal W}_{1}}{\partial
\phi^{\nu}}+{\cal K}_{1\nu\eta}\phi^{\eta}\frac{{\cal W}_{1}}{M^{2}}\right)-3\frac{|{\cal W}_{1}|^{2}}{M^{2}}\right]
   \end{array}\ee
with the general {\it K$\ddot{a}$hler metric}
 ${\cal K}_{1}^{\alpha\beta}=\frac{\partial^{2}{\cal K}_{1}}{\partial\phi_{\alpha}\partial\phi^{\dag}_{\beta}}$. In this context $A(13)$
factor comes as an outcome of dimension reduction and is explicitly mentioned in the appendix B.
In most of the simple situations, we are interested in the {\it Canonical metric} structure defined by ${\cal K}_{1}^{\alpha\beta}=\delta^{\alpha\beta}$.
Consequently the  ${\cal N}=1,{\cal D}=4$ SUGRA action turns out to be
\be\begin{array}{lllll}\label{ast5}\displaystyle S^{(4)}_{Can~Sugra}=
\frac{1}{2\kappa^{2}_{(4)}}\int d^{4}x \sqrt{-g^{(4)}}\left[M(T,T^{\dag})
g^{\alpha\beta (4)}(\partial_{\alpha}\phi_{\mu})^{\dag}(\partial_{\beta}\phi^{\mu})
-Z(T,T^{\dag})V^{(4)}_{Can}(\phi)
\right].\end{array}\ee
where the canonical F-term potential 
 can be recast as \be\label{totalpot}
V^{(4)}=V^{(4)}_{F}={\cal A}(13)
\exp\left(\frac{\phi^{\dagger}_{\alpha}\phi^{\alpha}}{M^{2}}\right)\left[\left|\frac{\partial
{\cal W}_{1}}{\partial \phi_{\beta}}\right|^{2}-3\frac{|{\cal W}_{1}|^{2}}{M^{2}}\right].\ee
To derive the expression for the specific form of the inflaton potential we start with a specific superpotential \cite{maya}
${\cal W}_{1}=v\phi-\frac{g}{n+1}\phi^{n+1}$,
 where $\phi$ is the {\it chiral} superfield with ${\it R}$ charge
$\frac{2}{n+1}$ introduced as an inflaton with $n\ge 2$. Here $g(\sim {\cal O}(1))$ is the positive and real coupling constant and $v$
is the VEV of the  field $\phi$. In this model $\it U(1)_{R}$ symmetry is dynamically broken to a discrete
 ${\it Z_{2n}R}$ symmetry at the scale $v<<1$. 
Consequently the inflaton transforms as $\phi\longrightarrow\exp\left(\frac{2\alpha i}{n+1}\right)\phi(\exp(-i\alpha)\theta)$, where 
$\alpha$ and $\theta$ are the local gauge parameters of supergravity theory.
This leads to the following form of the bulk contribution to the  potential 
\be\label{yu}
  V^{(4)}_{bulk}(\phi) ={\cal A}(13) \exp\left(|\phi|^2\right) \left[
              \left| \left( 1+|\phi|^2 \right)v^2
             -\left( 1+\frac{|\phi|^2}{n+1} \right) g\phi^n \right|^2 
             - 3 |\phi|^2 \left| v^2 - \frac{g}{n+1}\phi^n \right|^2\right].\ee

It has a minimum at
$|\phi|_{\rm min} \simeq \left(\frac{v^2}{g}\right)^{\frac{1}{n}}
{\rm and~~Im~}\phi^n_{\rm min}=0$ 
with negative energy density
\be\label{tresa}V ( \phi_{\rm min} ) \simeq 
            - 3{\cal A}(13)\exp\left(|\phi|^2\right) \left|{\cal W}_{1} ( \phi_{\rm min} ) \right|^2
                            \simeq 
            -3 {\cal A}(13)\left( \frac{n}{n+1} \right)^{2} v^{4} \left| \phi_{\rm min} \right|^2.\ee
 However, in
the context of SUGRA, we may interpret that such negative potential
energy is almost canceled by positive contribution due to the local
supersymmetry breaking, $\Lambda_{\rm SUGRA}^4$, and that the residual
positive energy density is responsible for the present dark
energy. Then, we can relate the energy scale of this model to the
gravitino mass as
\be\label{gmass} m_{3/2} \simeq \frac{n}{n+1} \left( \frac{v^2}{g} \right)^{\frac{1}{n}} v^2.\ee
Identifying the real part of $\phi$ with the inflaton $\phi\rightarrow
\sqrt{2} {\rm Re} \phi$, the dynamics of the inflaton is governed by the following potential,
\be\label{potential}
 V^{(4)}_{bulk}(\phi) \simeq {\cal A}(13)\left(v^4 - \frac{2g}{2^{n/2}}v^2 \phi^n 
  + \frac{g^2}{2^n}\phi^{2n}\right). 
\ee
Imposing renormalization condition, here we restrict ourselves to $n=2$ leading to effective ${\cal N}=1, {\cal D}=4$ SUGRA
potential
\be\label{pot1}
V^{(4)}_{bulk}(\phi) = {\cal A}(13)\left(v^4 - gv^2 \phi^2
  + \frac{g^2}{4}\phi^{4}\right). 
\ee

\section*{\bf Appendix B}
\label{apb1}

The effective action (\ref{model1}) leads to 4D effective {\it Einstein-Gauss Bonnet} equation as follows:
\be\label{en4}
G^{(4)}_{\alpha\beta}+\alpha^{}_{(4)}H^{(4)}_{\alpha\beta}
=8\pi G^{}_{(4)}T^{(4)}_{\alpha\beta}-\Lambda^{}_{(4)}g^{(4)}_{\alpha\beta}
\ee
where the  {\it Energy Momentum tensor}, {\it Einstein tensor} and {\it Gauss-Bonnet tensor} are given by
\be\begin{array}{llllllll}\label{emeff}
    T^{(4)}_{\alpha\beta}=T^{(bulk)(4)}_{\alpha\beta}+T^{(brane)(4)}_{\alpha\beta}
=-2X-\frac{{\cal S}_{(1)}R}{b_{0}A_{1}}g^{(4)}\left\{-Xg^{\rho\sigma(4)}+\frac{e^{5}_{\dot{5}}V^{(4)}_{Can}}{
16\pi^{2}R^{4}\beta^{2}}\right\}\\~~~~~~~~~~~~~~~~~~~~~~~~~~~~~~~~~~~~~~~~~~~~~~~~~~~~~~~~~~
+\hat{\tilde{K}}_{X}\nabla_{\alpha}\phi\nabla_{\beta}\phi+\tilde{K}g_{\alpha\beta}-2\nabla_{(\alpha}\tilde{G}_{\beta)}\phi
+g_{\alpha\beta}\nabla_{\lambda}\tilde{G}\nabla^{\lambda}\phi\\
~~~~~~~~~~~~~~~~~~~~~~~~~~~~~~~~~~~~~~~~~~~~~~~~~~~~~~~~~~~~~~~~~~~~~~~~~~~~~~~
-\tilde{G}_{X}\Box\phi\nabla_{\alpha}\phi\nabla_{\beta}\phi
   \end{array}\ee
\be\begin{array}{lllllll}\label{efen}
    G^{(4)}_{\alpha\beta}=\tilde{h}_{1}R^{(4)}_{\alpha\beta}-\frac{1}{2}\tilde{ h}_{2}g^{(4)}_{\alpha\beta}R_{(4)}
g^{(4)}_{\alpha\beta}\left\{6\tilde{ h}_{4}+4\tilde{h}_{5}-3\tilde{h}_{3}\right\}
   \end{array}\ee
\be\begin{array}{llllllll}\label{gbef4}
    H^{(4)}_{\alpha\beta}=\left\{R^{(4)}_{\alpha\rho\gamma\delta}R^{(4)}_{\beta\lambda\sigma\xi}\tilde{h}_{7}
+\frac{\tilde{h}_{8}}{R^{2}\beta^{2}}\left[R^{(4)}_{\beta\lambda\sigma\xi}
\left(g^{(4)}_{\gamma\rho}g^{(4)}_{\alpha\delta}-g^{(4)}_{\alpha\gamma}g^{(4)}_{\delta\rho}\right)
+R^{(4)}_{\beta\lambda\sigma\xi}
\left(g^{(4)}_{\gamma\rho}g^{(4)}_{\alpha\delta}-g^{(4)}_{\alpha\gamma}g^{(4)}_{\delta\rho}\right)\right]
+\right.\\ \left.~~~~~~~~~~~~~~\frac{1}{R^{2}\beta^{2}}\left(g^{(4)}_{\gamma\rho}g^{(4)}_{\alpha\delta}-g^{(4)}_{\alpha\gamma}g^{(4)}_{\delta\rho}\right)
\left(g^{(4)}_{\gamma\rho}g^{(4)}_{\alpha\delta}-g^{(4)}_{\alpha\gamma}g^{(4)}_{\delta\rho}\right)\right\}g^{\rho\lambda(4)}
g^{\gamma\sigma(4)}g^{\delta\xi(4)}-\left\{R^{(4)}_{\alpha\lambda\beta\sigma}R^{(4)}_{\sigma^{'}\rho}\tilde{h}_{11}\right.\\ \left.
~~~~~~~~~~~~+\frac{\tilde{h}_{12}}{R^{2}\beta^{2}}R^{(4)}_{\sigma^{'}\rho}\left(g^{(4)}_{\lambda\beta}g^{(4)}_{\alpha\sigma}
-g^{(4)}_{\alpha\beta}g^{(4)}_{\sigma\lambda}\right)-\frac{3\tilde{h}_{13}}{R^{2}\beta^{2}}g^{(4)}_{\sigma^{'}\rho}
-\frac{3\tilde{h}_{14}}{R^{4}\beta^{4}}g^{(4)}_{\sigma^{'}\rho}\left(g^{(4)}_{\lambda\beta}g^{(4)}_{\alpha\sigma}-g^{(4)}_{\alpha\beta}g^{(4)}_{\sigma\lambda}
\right)\right\}g^{\lambda\sigma^{'}(4)}g^{\sigma\rho(4)}\\
~~~~~~~~~~~~~-\left\{R^{(4)}_{\alpha\gamma}R^{(4)}_{\beta\delta}\tilde{h}_{16}
+\frac{9\tilde{h}_{17}}{\beta^{4}R^{4}}g^{(4)}_{\alpha\gamma}g^{(4)}_{\beta\delta}
-\frac{3\left(\tilde{h}_{18}+\tilde{h}_{19}\right)}{R^{2}\beta^{2}}\right\}g^{\gamma\delta(4)}
+R_{(4)}R^{(4)}_{\alpha\beta}\tilde{h}_{21}-\frac{1}{R^{2}\beta^{2}}\left\{3R_{(4)}g^{(4)}_{\alpha\beta}\tilde{h}_{22}\right. \\ \left.~~~~~~~~~~~~~~
+12R^{(4)}_{\alpha\beta}\left(\tilde{h}_{23}+\tilde{h}_{25}\right)
+\frac{36\left(\tilde{h}_{24}+\tilde{\hat{h}}_{25}\right)}{R^{4}\beta^{4}}g^{(4)}_{\alpha\beta}\right\}
-\frac{1}{2}g^{(4)}_{\alpha\beta}\left[\left\{R^{(4)}_{\rho\sigma\lambda\eta}R^{(4)}_{\alpha\beta\gamma\delta}\tilde{h}_{26}
+\frac{R^{(4)}_{\alpha\beta\gamma\delta}}{R^{2}\beta^{2}}\tilde{h}_{27}\left(g^{(4)}_{\lambda\sigma}g^{(4)}_{\rho\eta}
-g^{(4)}_{\rho\lambda}g^{(4)}_{\eta\sigma}\right)\right.\right.\\ \left.\left.~~~~~~~~~~~~~~
+\frac{R^{(4)}_{\rho\sigma\lambda\eta}}{R^{2}\beta^{2}}\tilde{h}_{28}\left(g^{(4)}_{\gamma\beta}g^{(4)}_{\alpha\delta}
-g^{(4)}_{\alpha\gamma}g^{(4)}_{\delta\beta}\right)+\frac{\tilde{h}_{29}}{R^{4}\beta^{4}}
\left(g^{(4)}_{\lambda\sigma}g^{(4)}_{\rho\eta}
-g^{(4)}_{\rho\lambda}g^{(4)}_{\eta\sigma}\right)\left(g^{(4)}_{\gamma\beta}g^{(4)}_{\alpha\delta}
-g^{(4)}_{\alpha\gamma}g^{(4)}_{\delta\beta}\right)\right\}g^{\rho\alpha(4)}g^{\sigma\beta(4)}g^{\gamma\lambda(4)}g^{\delta\eta(4)}
\right.\\ \left.~~~~~~~~~~~~-4g^{\alpha\sigma(4)}g^{\beta\rho(4)}\left\{R^{(4)}_{\sigma\rho}R^{(4)}_{\alpha\beta}\tilde{h}_{31}
-\frac{3}{R^{2}\beta^{2}}\left(R^{(4)}_{\alpha\beta}g^{(4)}_{\sigma\rho}\tilde{h}_{32}+R^{(4)}_{\sigma\rho}g^{(4)}_{\alpha\beta}
\tilde{h}_{33}\right)+\frac{9}{R^{4}\beta^{4}}g^{(4)}_{\sigma\rho}g^{(4)}_{\alpha\beta}\right\}
+\frac{1}{R^{4}\beta^{4}}\left(144\tilde{h}_{40}\right.\right.\\ \left.\left.~~~~~~~~~~~~~~~~
-64\tilde{h}_{36}+192\tilde{h}_{41}+64\tilde{h}_{42}\right)-\frac{1}{R^{2}\beta^{2}}R_{(4)}
\left(24\tilde{h}_{38}+16\tilde{h}_{39}\right)+\tilde{h}_{37}\left(R_{(4)}\right)^{2}\right]
\end{array}\ee

From FLRW metric in 4D, equation(\ref{en4}) results in the following
 {\it Friedmann equations}:
\be\begin{array}{llllllll}\label{frd1}
    H^{4}\tilde{g}_{1}+H^{2}\tilde{g}_{2}
+\dot{H}H^{2}\tilde{g}_{3}+\dot{H}^{2}\tilde{g}_{4}+\dot{H}\tilde{g}_{5}=\tilde{\Lambda}^{}_{(4)}+8\pi G^{}_{(4)}\rho^{}
   \end{array}\ee
 \be\begin{array}{llllllll}\label{frd1}
     H^{4}\tilde{f}_{1}+H^{2}\tilde{f}_{2}
+\dot{H}H^{2}\tilde{f}_{3}+\dot{H}^{2}\tilde{f}_{4}+\dot{H}\tilde{f}_{5}
=-\tilde{\hat{\Lambda}}^{}_{(4)}+8\pi G^{}_{(4)}p^{}
   \end{array}\ee
where the constants have been listed in appendix.
The {\it energy density} and {\it pressure} are now given by:
\be\label{effro}
\rho^{}=\left[2\tilde{K}_{X}X-\tilde{K}+6H\tilde{G}_{X}\dot{\phi}X-2\tilde{G}_{\phi}X-2X\left(1-\Theta_{1}\right)
+\Theta_{3}(T,T^{\dag})V_{bulk}(\phi)\right],\ee
\be\label{effp}
p^{}=\left[\tilde{K}-2\left(\tilde{G}_{\phi}+\tilde{G}_{X}\ddot{\phi}\right)X-2X\Theta_{1}-
\Theta_{3}\left(T,T^{\dag}\right)V_{bulk}(\phi)\right],\ee
where $\Theta_{1}=\frac{{\cal S}_{1}R}{b_{0}A_{1}}$ and $\Theta_{3}(T,T^{\dag})=\frac{\sqrt{2}}{16\pi^{2}b_{0}R^{3}(T+T^{\dag})}$.
All the constants appearing in this section are explicitly mentioned in the Appendix C.

\section*{\bf Appendix C}
\label{apb2}

In this section we have explicitly mentioned the expressions for the constants appearing in the section III, section IV and appendix A.

\be\begin{array}{llll}\label{aw1}
\displaystyle{\cal I}(1)
=\int^{+\pi R}_{-\pi R}dy\frac{4\left(3\exp(2\beta y)
+3T^{2}_{4}\exp(-2\beta y)-2T_{4}\right)}{R^{2}\left(\exp(\beta y)
+T_{4}\exp(-\beta y)\right)^{2}},
\end{array}\ee
\be\begin{array}{llll}\label{aw2}
\displaystyle{\cal C}(1)
=\frac{R^{7}}{b^{7}_{0}}\int^{+\pi R}_{-\pi R}dy\left(\exp(\beta y)+T_{4}\exp(-\beta y)\right)^{\frac{7}{2}},
\end{array}\ee
\be\begin{array}{llll}\label{aw3}
\displaystyle{\cal I}(2)={\cal A}(6)=\tilde{h}_{11}
=\frac{R^{3}}{b^{3}_{0}}\int^{+\pi R}_{-\pi R}dy\left(\exp(\beta y)+T_{4}\exp(-\beta y)\right)^{\frac{3}{2}},
\end{array}\ee
\be\begin{array}{llll}\label{aw4}
\displaystyle{\cal C}(2)
=\frac{\beta^{2}R^{5}}{4b^{5}_{0}}\int^{+\pi R}_{-\pi R}dy\frac{\left(\exp(\beta y)
+T_{4}\exp(-\beta y)\right)^{\frac{5}{2}}\left(T_{4}\exp(-\beta y)-\exp(\beta y)\right)^{2}}{\left(\exp(\beta y)
+T_{4}\exp(-\beta y)\right)^{2}},
\end{array}\ee
\be\begin{array}{llll}\label{aw5}
\displaystyle{\cal C}(4)={\cal I}(4)={\cal A}(5)
=\frac{\beta^{4}R^{3}}{16b^{3}_{0}}\int^{+\pi R}_{-\pi R}dy\frac{\left(\exp(\beta y)
+T_{4}\exp(-\beta y)\right)^{\frac{3}{2}}\left(T_{4}\exp(-\beta y)-\exp(\beta y)\right)^{4}}{\left(\exp(\beta y)
+T_{4}\exp(-\beta y)\right)^{4}},
\end{array}\ee
\be\begin{array}{llll}\label{aw6}
\displaystyle{\cal A}(7)
=\frac{\beta^{4}b^{5}_{0}}{16R^{5}}\int^{+\pi R}_{-\pi R}dy\frac{\left(T_{4}\exp(-\beta y)-\exp(\beta y)\right)^{4}}
{\left(\exp(\beta y)
+T_{4}\exp(-\beta y)\right)^{4}\left(\exp(\beta y)
+T_{4}\exp(-\beta y)\right)^{\frac{5}{2}}},
\end{array}\ee
\be\begin{array}{llll}\label{aw7}
\displaystyle{\cal A}(8)
=\frac{4\beta^{4}b^{5}_{0}T^{2}_{4}}{R^{5}}\int^{+\pi R}_{-\pi R}dy\frac{1}
{\left(\exp(\beta y)
+T_{4}\exp(-\beta y)\right)^{4}\left(\exp(\beta y)
+T_{4}\exp(-\beta y)\right)^{\frac{5}{2}}},
\end{array}\ee
\be\begin{array}{llll}\label{aw8}
\displaystyle{\cal A}(9)=\tilde{h}_{4}=\tilde{h}_{30}
=\frac{\beta^{2}b^{5}_{0}}{4R^{5}}\int^{+\pi R}_{-\pi R}dy\frac{\left(T_{4}\exp(-\beta y)-\exp(\beta y)\right)^{2}}
{\left(\exp(\beta y)
+T_{4}\exp(-\beta y)\right)^{2}\left(\exp(\beta y)
+T_{4}\exp(-\beta y)\right)^{\frac{5}{2}}}
\end{array}\ee
\be\begin{array}{llll}\label{aw9}
\displaystyle{\cal A}(10)=\tilde{h}_{5}
=-\frac{2\beta^{2}b^{5}_{0}T_{4}}{R^{5}}\int^{+\pi R}_{-\pi R}dy\frac{1}
{\left(\exp(\beta y)
+T_{4}\exp(-\beta y)\right)^{2}\left(\exp(\beta y)
+T_{4}\exp(-\beta y)\right)^{\frac{5}{2}}},
\end{array}\ee
\be\begin{array}{llll}\label{aw10}
\displaystyle{\cal A}(11)
=-\frac{2\beta^{4}b^{5}_{0}T_{4}}{2R^{5}}\int^{+\pi R}_{-\pi R}dy\frac{\left(T_{4}\exp(-\beta y)-\exp(\beta y)\right)^{2}}
{\left(\exp(\beta y)
+T_{4}\exp(-\beta y)\right)^{4}\left(\exp(\beta y)
+T_{4}\exp(-\beta y)\right)^{\frac{5}{2}}},
\end{array}\ee
\be\begin{array}{llll}\label{aw11}
\displaystyle{\cal A}(12)
=-\frac{b_{0}}{R}\int^{+\pi R}_{-\pi R}dy\frac{\exp\left(-\frac{8\pi i y}{R}\right)}
{\sqrt{\left(\exp(\beta y)
+T_{4}\exp(-\beta y)\right)}},
\end{array}\ee
\be\begin{array}{llll}\label{aw12}
\displaystyle{\cal A}(1)=\tilde{h}_{1}
=\frac{b_{0}}{R}\int^{+\pi R}_{-\pi R}dy\frac{1}
{\sqrt{\left(\exp(\beta y)
+T_{4}\exp(-\beta y)\right)}},
\end{array}\ee

\be\begin{array}{llll}\label{aw14}
  \displaystyle {\cal A}(13)=\frac{b_{0}\tilde{h}_{16}}{R}=\frac{b_{0}\tilde{h}_{31}}{R}=\int^{+\pi R}_{-\pi R}dy \sqrt{\exp(\beta y)+T_{4}\exp(-\beta y)},\\
\end{array}\ee
\be\begin{array}{llll}\label{aw15}
 \displaystyle   \tilde{h}_{2}=\frac{b^{5}_{0}}{R^{5}}\int^{+\pi R}_{-\pi R}dy\frac{1}
{\left(\exp(\beta y)
+T_{4}\exp(-\beta y)\right)^{\frac{5}{2}}},\\
   \end{array}\ee
\be\begin{array}{llll}\label{aw16}
 \displaystyle   \tilde{h}_{3}=\frac{b_{0}\beta^{2}}{4R}\int^{+\pi R}_{-\pi R}dy\frac{\left(T_{4}\exp(-\beta y)-\exp(-\beta y)\right)^{2}}
{\left(\exp(\beta y)
+T_{4}\exp(-\beta y)\right)^{\frac{5}{2}}},\\
   \end{array}\ee
\be\begin{array}{llll}\label{aw17}
 \displaystyle   \tilde{h}_{6}=\tilde{h}_{21}=\frac{b^{3}_{0}}{R^{3}}\int^{+\pi R}_{-\pi R}dy\frac{1}
{\left(\exp(\beta y)
+T_{4}\exp(-\beta y)\right)^{\frac{3}{2}}},\\
   \end{array}\ee
\be\begin{array}{llll}\label{aw18}
 \displaystyle   \tilde{h}_{7}=\tilde{h}_{26}=\frac{R^{5}}{b^{5}_{0}}\int^{+\pi R}_{-\pi R}dy
\left(\exp(\beta y)
+T_{4}\exp(-\beta y)\right)^{\frac{5}{2}},\\
   \end{array}\ee
\be\begin{array}{llll}\label{aw19}
 \displaystyle   \tilde{h}_{8}=\tilde{h}_{13}=\tilde{h}_{27}=\tilde{h}_{28}=\frac{\beta^{2}R^{3}}{4b^{3}_{0}}\int^{+\pi R}_{-\pi R}dy\frac{\left(T_{4}\exp(-\beta y)
-\exp(\beta y)\right)^{2}}
{\left(\exp(\beta y)
+T_{4}\exp(-\beta y)\right)^{\frac{1}{2}}},\\
   \end{array}\ee
\be\begin{array}{llll}\label{aw20}
 \displaystyle   \tilde{h}_{9}=\tilde{h}_{14}=\tilde{h}_{17}=\tilde{h}_{29}=\tilde{h}_{34}=\tilde{h}_{36}=\frac{\beta^{4}R}{16b_{0}}\int^{+\pi R}_{-\pi R}dy\frac{\left(T_{4}\exp(-\beta y)
-\exp(\beta y)\right)^{4}}
{\left(\exp(\beta y)
+T_{4}\exp(-\beta y)\right)^{\frac{7}{2}}},\\
   \end{array}\ee
\be\begin{array}{llll}\label{aw21}
 \displaystyle   \tilde{h}_{10}=\tilde{h}_{12}=\tilde{h}_{15}=\tilde{h}_{18}=\tilde{h}_{19}=\tilde{h}_{20}=
\tilde{h}_{32}=\tilde{h}_{33}=\tilde{h}_{35}=\frac{\beta^{2}R}{4b_{0}}\int^{+\pi R}_{-\pi R}dy\frac{\left(T_{4}\exp(-\beta y)
-\exp(\beta y)\right)^{2}}
{\left(\exp(\beta y)
+T_{4}\exp(-\beta y)\right)^{\frac{3}{2}}},\\
   \end{array}\ee
\be\begin{array}{llll}\label{aw22}
 \displaystyle   \tilde{h}_{22}=\tilde{h}_{23}=\frac{\beta^{2}b^{3}_{0}}{4R^{3}}\int^{+\pi R}_{-\pi R}dy\frac{\left(T_{4}\exp(-\beta y)
-\exp(\beta y)\right)^{2}}
{\left(\exp(\beta y)
+T_{4}\exp(-\beta y)\right)^{\frac{7}{2}}},\\
   \end{array}\ee
\be\begin{array}{llll}\label{aw23}
 \displaystyle   \tilde{h}_{24}=\frac{\beta^{4}b^{3}_{0}}{64R^{3}}\int^{+\pi R}_{-\pi R}dy\frac{\left(T_{4}\exp(-\beta y)
-\exp(\beta y)\right)^{4}}
{\left(\exp(\beta y)
+T_{4}\exp(-\beta y)\right)^{\frac{11}{2}}},\\
   \end{array}\ee
\be\begin{array}{llll}\label{aw24}
 \displaystyle   \tilde{h}_{25}=-\frac{2\beta^{2}T_{4}b^{3}_{0}}{R^{3}}\int^{+\pi R}_{-\pi R}dy\frac{1}
{\left(\exp(\beta y)
+T_{4}\exp(-\beta y)\right)^{\frac{7}{2}}},\\
   \end{array}\ee
\be\begin{array}{llll}\label{aw241}
 \displaystyle   \tilde{\hat{h}}_{25}=-\frac{\beta^{4}T_{4}b^{3}_{0}}{2R^{3}}\int^{+\pi R}_{-\pi R}dy\frac{\left(T_{4}\exp(-\beta y)
-\exp(\beta y)\right)^{2}}
{\left(\exp(\beta y)
+T_{4}\exp(-\beta y)\right)^{\frac{11}{2}}},\\
   \end{array}\ee
\be\begin{array}{llll}\label{aw25}
 \displaystyle   \tilde{h}_{37}=\frac{b^{7}_{0}}{R^{7}}\int^{+\pi R}_{-\pi R}dy\frac{1}
{\left(\exp(\beta y)
+T_{4}\exp(-\beta y)\right)^{\frac{7}{2}}},\\
   \end{array}\ee
\be\begin{array}{llll}\label{aw26}
 \displaystyle   \tilde{h}_{38}=\frac{\beta^{2}b^{7}_{0}}{4R^{7}}\int^{+\pi R}_{-\pi R}dy\frac{\left(T_{4}\exp(-\beta y)
-\exp(\beta y)\right)^{2}}
{\left(\exp(\beta y)
+T_{4}\exp(-\beta y)\right)^{\frac{11}{2}}},\\
   \end{array}\ee
\be\begin{array}{llll}\label{aw27}
 \displaystyle   \tilde{h}_{39}=-\frac{2\beta^{2}T_{4}b^{7}_{0}}{R^{7}}\int^{+\pi R}_{-\pi R}dy\frac{1}
{\left(\exp(\beta y)
+T_{4}\exp(-\beta y)\right)^{\frac{11}{2}}},\\
   \end{array}\ee
\be\begin{array}{llll}\label{aw28}
 \displaystyle   \tilde{h}_{40}=\frac{\beta^{4}b^{7}_{0}}{16R^{7}}\int^{+\pi R}_{-\pi R}dy\frac{\left(T_{4}\exp(-\beta y)
-\exp(\beta y)\right)^{4}}
{\left(\exp(\beta y)
+T_{4}\exp(-\beta y)\right)^{\frac{15}{2}}},\\
   \end{array}\ee
\be\begin{array}{llll}\label{aw29}
 \displaystyle   \tilde{h}_{41}=-\frac{\beta^{4}T_{4}b^{7}_{0}}{2R^{7}}\int^{+\pi R}_{-\pi R}dy\frac{\left(T_{4}\exp(-\beta y)
-\exp(\beta y)\right)^{2}}
{\left(\exp(\beta y)
+T_{4}\exp(-\beta y)\right)^{\frac{15}{2}}},\\
   \end{array}\ee
\be\begin{array}{llll}\label{aw30}
 \displaystyle   \tilde{h}_{42}=\frac{4\beta^{4}T^{2}_{4}b^{7}_{0}}{R^{7}}\int^{+\pi R}_{-\pi R}dy\frac{1}
{\left(\exp(\beta y)
+T_{4}\exp(-\beta y)\right)^{\frac{15}{2}}},\\
   \end{array}\ee

\be\begin{array}{lllllllllllllllllllll}\label{const}
\displaystyle\tilde{g}_{1}=\tilde{\alpha}_{1}+\tilde{\alpha}_{4}+\tilde{\alpha}_{5}+a^{2}\left(\tilde{\alpha}_{6}+\tilde{\alpha}_{7}\right),~~
\tilde{g}_{2}=\left(\tilde{\alpha}_{2}+\tilde{\alpha}_{3}\right), ~~\tilde{g}_{3}=2\tilde{\alpha}_{1}+\tilde{\alpha}_{5}+
a^{2}\left(\tilde{\alpha}_{6}+2\tilde{\alpha}_{7}\right),\\
\tilde{g}_{4}=\tilde{\alpha}_{1}+a^{2}\tilde{\alpha}_{7},~~ \tilde{g}_{5}=\tilde{\alpha}_{2},\\
\tilde{f}_{1}=\tilde{\beta}_{1}+\tilde{\beta}_{5}+\tilde{\beta}_{6},~~\tilde{f}_{2}=\tilde{\beta}_{2}+\tilde{\beta}_{3},~~
\tilde{f}_{3}=2\tilde{\beta}_{1}+\tilde{\beta}_{6},~~\tilde{f}_{4}=\tilde{\beta}_{1},~~\tilde{f}_{5}=\tilde{\beta}_{3},\\
\tilde{\alpha}_{1}=\alpha_{(4)}\left(-3\tilde{h}_{7}-9\tilde{h}_{16}-18\tilde{h}_{21}+\frac{3}{2}\tilde{h}_{26}
-18\tilde{h}_{31}+36\tilde{h}_{39}\right),\\
\tilde{\alpha}_{2}=3\left(\tilde{h}_{2}-\tilde{h}_{1}\right)+\frac{\alpha_{(4)}}{R^{2}\beta^{2}}
\left(12\tilde{h}_{8}-3\tilde{h}_{12}+9\tilde{h}_{13}+9\tilde{h}_{18}+9\tilde{h}_{19}
+18\tilde{h}_{22}+36\tilde{h}_{23}\right.\\ \left.
~~~~~~~~~~~~~~~~~~~~~~~~~~~~~~~~~~~~~~~~~~~~~~~~~~~~~~~+24\tilde{h}_{25}
+18\tilde{h}_{32}+18\tilde{h}_{33}-72\tilde{h}_{38}-48\tilde{h}_{39}\right),\\
\tilde{\alpha}_{3}=3\tilde{h}_{2}-\frac{\alpha_{(4)}}{R^{2}\beta^{2}}
\left(18\tilde{h}_{22}-6\tilde{h}_{12}-72\tilde{h}_{38}-48\tilde{h}_{39}\right),~~~~~
\tilde{\alpha}_{4}=36\alpha_{(4)}\tilde{h}_{39},\\
\tilde{\alpha}_{5}=\alpha_{(4)}\left(72\tilde{h}_{39}-18\tilde{h}_{21}\right),~~~~~
\tilde{\alpha}_{6}=6\alpha_{(4)}\tilde{h}_{11},~~~~~\tilde{\alpha}_{7}=3\alpha_{(4)}\tilde{h}_{11},\\
\tilde{\Lambda}^{}_{4}=\Lambda^{}_{4}-\frac{\alpha_{(4)}}{R^{4}\beta^{4}}\left(32\tilde{h}_{42}
+96\tilde{h}_{41}+72\tilde{h}_{40}-64\tilde{h}_{36}-18\tilde{h}_{34}-24\tilde{\hat{h}}_{25}-36\tilde{h}_{24}
-9\tilde{h}_{17}-6\tilde{h}_{9}\right)-\frac{9\tilde{h}_{14}}{R^{2}\beta^{2}},\\
\tilde{\beta}_{1}=\alpha_{(4)}\left(\tilde{h}_{7}-3\tilde{h}_{11}-\tilde{h}_{16}
+6\tilde{h}_{21}-\frac{\tilde{h}_{26}}{2}+2\tilde{h}_{31}-3\tilde{h}_{37}\right),\\
\tilde{\beta}_{2}=\left(2\tilde{h}_{1}-3\tilde{h}_{2}\right)+\alpha_{(4)}\left\{\frac{1}{R^{2}\beta^{2}}
\left(-8\tilde{h}_{8}+4\tilde{h}_{12}+6\tilde{h}_{13}+6\tilde{h}_{18}+6\tilde{h}_{14}-18\tilde{h}_{22}
-24\tilde{h}_{23}\right.\right.\\ \left.\left.~~~~~~~~~~~~~~~~~~~~~~~~~~~~~~~~~~~~~~
-16\tilde{h}_{25}-12\tilde{h}_{32}-12\tilde{h}_{33}+72\tilde{h}_{38}+48\tilde{h}_{39}\right)-\tilde{h}_{26}\right\},\\
\tilde{\beta}_{3}=\left(\tilde{h}_{1}-3\tilde{h}_{2}\right)+\frac{\alpha_{(4)}}{R^{2}\beta^{2}}
\left(-4\tilde{h}_{8}+5\tilde{h}_{12}+3\tilde{h}_{13}+3\tilde{h}_{18}+3\tilde{h}_{19}-18\tilde{h}_{22}-12\tilde{h}_{23}
\right.\\ \left.~~~~~~~~~~~~~~~~~~~~~~~~~~~~~~~~~~~~~~~~~~~~~~~~~~~~~~~~~~~~~~~~~~~~~~~~
-8\tilde{h}_{25}-6\tilde{h}_{32}-6\tilde{h}_{33}+72\tilde{h}_{38}+48\tilde{h}_{39}\right),\\
\tilde{\beta}_{5}=\alpha_{(4)}\left(-3\tilde{h}_{37}+8\tilde{h}_{31}+12\tilde{h}_{21}
-4\tilde{h}_{16}-4\tilde{h}_{11}+2\tilde{h}_{7}\right),\\
\tilde{\beta}_{6}=\alpha_{(4)}\left(-3\tilde{h}_{37}+8\tilde{h}_{31}+18\tilde{h}_{21}
-4\tilde{h}_{16}-2\tilde{h}_{11}\right),\\
\tilde{\hat{\Lambda}}^{}_{4}=\left(6\tilde{h}_{4}+4\tilde{h}_{5}-3\tilde{h}_{3}+\Lambda^{}_{4}\right)
+\frac{\alpha_{(4)}}{R^{4}\beta^{4}}\left(72\tilde{h}_{40}+96\tilde{h}_{41}+64\tilde{h}_{42}
-64\tilde{h}_{36}+9\tilde{h}_{34}\right.\\ \left.~~~~~~~~~~~~~~~~~~~~~~~~~~~~~~~~~~
~~~~~~~~~~~~~~~~~~~~~~~~~+24\tilde{\hat{h}}_{25}+36\tilde{h}_{24}-9\tilde{h}_{17}-9\tilde{h}_{14}+6\tilde{h}_{9}\right).
\end{array}\ee
\be\begin{array}{llll}\label{aw13}
  \displaystyle Q_{1}=\frac{{\cal I}(3)}{\left(\bar{C}_{1}d_{2}-\bar{C}_{2}d_{4}\right)},
~~Q=\frac{\left(\bar{C}_{2}d_{3}-\bar{C}_{1}d_{1}\right)}{\left(\bar{C}_{1}d_{2}-\bar{C}_{2}d_{4}\right)},\\
   \displaystyle \bar{C}_{1}=\frac{b_{0}\exp(-2\beta\pi R)}
{2R\beta\left[\left(\frac{4\pi}{R}\right)+i\beta\right]
\left[\left(\frac{4\pi}{R}\right)+5i\beta\right]T^{\frac{3}{2}}_{4}}
\sqrt{\frac{\exp(-2\beta \pi R)+T_{4}}{\exp(-\beta\pi R)+T_{4}\exp(\beta\pi R)}},\\
\displaystyle \bar{C}_{2}=\frac{b_{0}\exp(-4i\beta\pi^{2})}
{2R\beta\left[\left(\frac{4\pi}{R}\right)+i\beta\right]
\left[\left(\frac{4\pi}{R}\right)+5i\beta\right]T^{\frac{3}{2}}_{4}}
\sqrt{\frac{\exp(2\beta \pi R)+T_{4}}{\exp(\beta\pi R)+T_{4}\exp(-\beta\pi R)}},\\
\displaystyle{\cal I}(3)
=\frac{b_{0}}{R}\int^{+\pi R}_{-\pi R}dy\frac{\exp\left(-\frac{2\pi i y}{R}\right)}
{\sqrt{\left(\exp(\beta y)
+T_{4}\exp(-\beta y)\right)}},\end{array}\ee

\be\begin{array}{llll}\label{laskta}\displaystyle d_{2}=6\beta\exp(2i\pi^{2})\left[
\left(\beta-\frac{4i\pi}{R}\right)~_2F_1\left[\frac{3}{2};\frac{5}{4}-\frac{\pi i}{R\beta};
\frac{9}{4}-\frac{\pi i}{R\beta};-\frac{\exp(-2\beta\pi R)}{T_{(4)}}\right]
\right.\\ \left.~~~~~~~\displaystyle+\left(5\beta-\frac{4i\pi}{R}\right)T_{4}\exp(2\beta\pi R)~_2F_1\left[\frac{3}{2};\frac{1}{4}-\frac{\pi i}{R\beta};
\frac{5}{4}-\frac{\pi i}{R\beta};-\frac{\exp(-2\beta\pi R)}{T_{(4)}}\right]
\right],\\
\displaystyle d_{4}=6\beta\left[
\left(\beta-\frac{4i\pi}{R}\right)~_2F_1\left[\frac{3}{2};\frac{5}{4}-\frac{\pi i}{R\beta};
\frac{9}{4}-\frac{\pi i}{R\beta};-\frac{\exp(2\beta\pi R)}{T_{(4)}}\right]
\right.\\ \left.~~~~~~~\displaystyle+\left(5\beta-\frac{4i\pi}{R}\right)T_{4}~_2F_1\left[\frac{3}{2};\frac{1}{4}-\frac{\pi i}{R\beta};
\frac{5}{4}-\frac{\pi i}{R\beta};-\frac{\exp(2\beta\pi R)}{T_{(4)}}\right]
\right],\\
\displaystyle d_{3}=\left\{\frac{16\pi^{2}}{R^{2}}+\frac{24\pi i \beta}
{R}-5\beta^{2}\right\}\exp\left[\left(\beta+\frac{2i\pi}{R}\right)\pi R\right]
~_2F_1\left[\frac{3}{4};\frac{3}{2};
\frac{7}{4};-\frac{\exp(2\beta\pi R)}{T_{(4)}}\right],\\
\displaystyle  d_{1}=\left\{\frac{16\pi^{2}}{R^{2}}+\frac{24\pi i \beta}
{R}-5\beta^{2}\right\}\exp(\beta\pi R)
~_2F_1\left[\frac{3}{4};\frac{3}{2};
-\frac{7}{4};-\frac{\exp(-2\beta\pi R)}{T_{(4)}}\right],
   \end{array}
\ee
\section*{\bf Appendix D}
The determinant appearing in the first term of the equation(\ref{kl}) can be written as
\be\begin{array}{llll}\label{det1h}\displaystyle {\cal D}=det\left(\delta^{A}_{B}+f(\Phi)G_{CD}\partial^{A}\Phi^{C}\partial_{B}\Phi^{D}+b_{CD}\partial^{A}\Phi^{C}\partial_{B}\Phi^{D}
+\Theta(\Phi)F^{A}_{B}\right)=det\left(I+{\cal S}+{\cal B}\right)\\
\displaystyle~~~=-\frac{1}{4!}\epsilon_{ABCD}\epsilon^{EFHJ}\left(I+{\cal S}+{\cal B}\right)^{A}_{E}\left(I+{\cal S}+{\cal B}\right)^{B}_{F}\left(I+{\cal S}+{\cal B}\right)^{C}_{H}
\left(I+{\cal S}+{\cal B}\right)^{D}_{J},\end{array}\ee
where we define 
\be\begin{array}{lllll}\label{cvcv}\displaystyle\Theta(\Phi)=2\pi\alpha^{'}\sqrt{h(y)},
~~~~~ \displaystyle{\cal S}^{A}_{B}=f(\phi)G_{AB}\partial^{A}\Phi^{C}\partial_{B}\Phi^{D},
\\~\displaystyle{\cal B}^{A}_{B}=b_{CD}\partial^{A}\Phi^{C}\partial_{B}\Phi^{D}
+\Theta(\Phi)F^{A}_{B}=\frac{\Theta(\Phi)}{2\pi\alpha^{'}}g^{AC}{\cal F}_{CB}.\end{array}\ee
In this connection $I$, ${\cal S}$ and ${\cal B}$ are all $5\times 5$ matrices satisfies the property ${\cal S}_{AB}={\cal S}_{BA}$, ${\cal B}_{AB}=-{\cal B}_{BA}$.
Moreover ${\cal F}_{CB}$ is a {\it Neveu-Schwarz} gauge invariant. Detailed computation of the determinant yields 
\be\begin{array}{llllll}\label{jkhgfg}\displaystyle {\cal D}=
{\cal D}_{{\cal S}}-\frac{1}{2}Tr({\cal B}^{2})\left(1+Tr({\cal S})\right)+Tr({\cal S}{\cal B}^{2})\left(1+Tr({\cal S})\right)-Tr({\cal S}^{2}{\cal B}^{2})\\~~~~~~~~~~~~~
\displaystyle-\frac{1}{4}Tr({\cal B}^{2})\left[\left(Tr({\cal S})\right)^{2}-Tr({\cal S}^{2})\right]
-\frac{1}{2}Tr({\cal S}{\cal B}{\cal S}{\cal B})+
\frac{1}{8}\left[\left(Tr({\cal B}^{2})\right)^{2}-2Tr({\cal S}^{4})\right]
\end{array}\ee
where
\be\begin{array}{llllll}\label{jkhgfgw}\displaystyle 
{\cal D}_{{\cal S}}=1+Tr({\cal S})+\frac{1}{2}\left[\left(Tr({\cal S})\right)^{2}-Tr({\cal S}^{2})\right]
+S_{A}^{[A}S_{B}^{B}S_{C}^{C]}++S_{A}^{[A}S_{B}^{B}S_{C}^{C}S_{D}^{D]}.
\end{array}\ee
Now throughout our discussion we assume that the D4 brane and bulk SUGRA form fields are ignored then we can write ${\cal D}\simeq {\cal D}_{{\cal S}}$.
In this connection we use the well known identity 
\be\begin{array}{lll}\label{mnbv}\displaystyle det(I+{\cal S})=
    -\frac{1}{4!}\epsilon_{ABCD}\epsilon^{EFHJ}\left(I+{\cal S}\right)^{A}_{E}\left(I+{\cal S}\right)^{B}_{F}\left(I+{\cal S}\right)^{C}_{H}
\left(I+{\cal S}\right)^{D}_{J}
=e^{Tr(\ln(I+{\cal S}))},\\
Tr({\cal S}^{n})=Tr((-2f(\Phi)X)^{n})
   \end{array}\ee
which results in
\be\begin{array}{llllll}\label{jazgw}\displaystyle
{\cal D}\simeq{\cal D}_{{\cal S}}=1-2f(\Phi)G_{AB}X^{AB}+4f^{2}(\Phi)X_{A}^{[A}X_{B}^{B]}
-8f^{3}(\Phi)X_{A}^{[A}X^{B}_{B}X^{C]}_{C}\\~~~~~~~~~~~~~~~~~~~~~~~~~~~~~~~~~~~~
~~~~~~~~~~~+16 f^{4}(\Phi)X_{A}^{[A}X^{B}_{B}X^{C}_{C}X^{D]}_{D}-32f^{5}(\Phi)X_{A}^{[A}X^{B}_{B}X^{C}_{C}X^{D}_{D}X^{E]}_{E}\end{array}\ee
where the brackets denote antisymmetrisation on the field indices.



\end{document}